\documentclass[fleqn,usenatbib]{mnras}

\usepackage{newtxtext,newtxmath}

\usepackage[T1]{fontenc}

\DeclareRobustCommand{\VAN}[3]{#2}
\let\VANthebibliography\thebibliography
\def\thebibliography{\DeclareRobustCommand{\VAN}[3]{##3}\VANthebibliography}

\usepackage{graphicx}	
\usepackage{amsmath}	
\usepackage{amssymb}	
\usepackage{xcolor}
\usepackage[export]{adjustbox}
\usepackage{flushend}
\usepackage{lastpage}
\usepackage[utf8]{inputenc}
\usepackage[normalem]{ulem}
\usepackage{comment}
\usepackage{url}
\usepackage{upgreek}

\usepackage[compact]{titlesec}  
\titlespacing{\section}{0pt}{12pt}{7pt}
\titlespacing{\subsection}{0pt}{9pt}{4pt}

\newcommand{\dmu}{\,pc\,cm$^{-3}$ }

\author[Main et al.]{R.~A.~Main$^1\thanks{Email: \href{mailto:ramain@mpifr-bonn.mpg.de}{ramain@mpifr-bonn.mpg.de}}$, J.~Antoniadis$^{2,3}$, S.~Chen$^{4}$, I.~Cognard$^{5,6}$, H.~Hu$^{1}$,  J.~Jang$^{1}$,  
R.~Karuppusamy$^{1}$, M.~Kramer$^{1, 7}$, \newauthor  K.~Liu$^{1}$, Y.~Liu$^{8,9}$, G.~Mall$^{10,11,1}$, J.~W.~McKee$^{12, 13}$,  M.~B.~Mickaliger$^{7}$,
D.~Perrodin$^{14}$, S.~A.~Sanidas$^{7}$, \newauthor B.~W.~Stappers$^{7}$, T.~Sprenger$^{1}$, O.~Wucknitz$^{1}$, C.~G.~Bassa$^{15}$, M.~Burgay$^{14}$, R.~Concu$^{14}$, M.~Gaikwad$^{1}$, \newauthor
G.~H.~Janssen$^{15, 16}$, K.~J.~Lee$^{4}$, A.~Melis$^{14}$, M.~Pilia$^{14}$, A.~Possenti$^{14, 17}$, L.~Wang$^{4}$, W.~W.~Zhu$^{18}$
  \\
 $^{1}$Max-Planck-Institut f\"{u}r Radioastronomie, Auf dem H\"{u}gel 69, 53121, Bonn, Germany \\
 $^{2}$Institute of Astrophysics, FORTH, Dept. of Physics, University of Crete, Voutes, University Campus, GR-71003 Heraklion, Greece \\
 $^{3}$Argelander Institut f\"{u}r Astronomie, Auf dem H\"{u}gel 71, 53121, Bonn, Germany \\
 $^{4}$Kavli Institute for Astronomy and Astrophysics, Peking University, Beijing 100871, P. R. China \\
 $^{5}$Laboratoire de Physique et Chimie de l'Environnement et de l'Espace LPC2E CNRS-Universit{\'e} d'Orl{\'e}ans, F-45071, Orl{\'e}ans, France \\
 $^{6}$Station de radioastronomie de Nan{\c c}ay, Observatoire de Paris, PSL Research University, CNRS/INSU F-18330 Nan{\c c}ay, France \\
 $^{7}$Jodrell Bank Centre for Astrophysics, School of Physics and Astronomy, The University of Manchester, Manchester M13 9PL, UK \\
 $^{8}${CAS Key Laboratory of FAST, National Astronomical Observatories, Chinese Academy of Sciences, Beijing 100101, China} \\
 $^{9}${Fakult\"at f\"ur Physik, Universit\"at Bielefeld, Postfach 100131, 33501 Bielefeld, Germany} \\
 $^{10}$Canadian Institute for Theoretical Astrophysics, University of Toronto, 60 St. George Street, Toronto, ON M5S 3H8, Canada \\
 $^{11}$Department of Physics, University of Toronto, 60 St. George Street, Toronto, ON M5S 1A7, Canada \\
 $^{12}${E.A. Milne Centre for Astrophysics, University of Hull, Cottingham Road, Kingston-upon-Hull, HU6 7RX, UK} \\
 $^{13}${Centre of Excellence for Data Science, Artificial Intelligence and Modelling (DAIM), University of Hull, Cottingham Road, Kingston-upon-Hull, HU6 7RX, UK} \\
 $^{14}$INAF - Osservatorio Astronomico di Cagliari, via della Scienza 5, I-09047 Selargius (CA), Italy \\
 $^{15}$ASTRON, the Netherlands Institute for Radio Astronomy, Oude Hoogeveensedijk 4, 7991 PD Dwingeloo, The Netherlands \\
 $^{16}$Department of Astrophysics/IMAPP, Radboud University, P.O. Box 9010, 6500 GL Nijmegen, The Netherlands \\
 $^{17}$Universit$\grave{a}$ di Cagliari, Dipartimento di Fisica, S.P. Monserrato-Sestu Km 0,700, I-09042 Monserrato (CA), Italy \\
 $^{18}$National Astronomical Observatories, Chinese Academy of Sciences, A20 Datun Rd, Chaoyang District, Beijing 100012, P.\,R.\,China \\
}

\title[Scintillation Arcs of MSPs with LEAP]{Variable Scintillation Arcs of Millisecond Pulsars observed with the Large European Array for Pulsars}

\date{Accepted XXX. Received YYY; in original form ZZZ}

\pubyear{2022}

\begin{document}
\label{firstpage}
\pagerange{\pageref{firstpage}--\pageref{lastpage}}
\maketitle

\begin{abstract}
We present the first large sample of scintillation arcs in millisecond pulsars, analysing 12 sources observed with the Large European Array for Pulsars (LEAP), and the Effelsberg 100\,m telescope. 
We estimate the delays from multipath propagation, measuring significant correlated changes in scattering timescales over a 10-year timespan. 
Many sources show compact concentrations of power in the secondary spectrum, which in PSRs J0613$-$0200 and J1600$-$3053 can be tracked between observations, and are consistent with compact scattering at fixed angular positions. 
Other sources such as PSRs J1643$-$1224 and J0621+1002 show diffuse, asymmetric arcs which are likely related to phase-gradients across the scattering screen.
PSR B1937+21 shows at least three distinct screens which dominate at different times 
and evidence of varying screen axes or multi-screen interactions.  
We model annual and orbital arc curvature variations in PSR J0613$-$0200, providing a measurement of the longitude of ascending node, resolving the sense of the orbital inclination, where our best fit model is of a screen with variable axis of anisotropy over time, corresponding to changes in the scattering of the source.  Unmodeled variations of the screen's axis of anisotropy are likely to be a limiting factor in determining orbital parameters with scintillation, requiring careful consideration of variable screen properties, or independent VLBI measurements.  Long-term scintillation studies such as this serve as a complementary tool to pulsar timing, to measure a source of correlated noise for pulsar timing arrays, solve pulsar orbits, and to understand the astrophysical origin of scattering screens.
\end{abstract}


\section{Introduction}

Pulsar Timing Arrays (PTAs) involve timing an ensemble of millisecond pulsars (MSPs) at different sky positions to detect nHz gravitational waves (GWs) from coalescing supermassive black holes.  Recently, PTAs have detected a common red-noise signal, which is a time-correlated signal of similar amplitude and spectrum shared among pulsars in the array \citep{arzoumanian+20, goncharov+21, chen+21, antoniadis+22}.  While it is possible that a gravitational wave signature is responsible for this effect, there is yet no detection of a spatial correlation that would be a smoking-gun of the gravitational wave background \citep{hellings+83}. 

The ionized interstellar medium (IISM) is one of the largest contributors of correlated noise to PTAs (see \citealt{verbiest+18} for a review), and understanding all of its effects is crucial, especially as a GW detection may be imminent. The total column density of electrons induces a $\lambda^2$ dispersive delay (where $\lambda$ is the observing wavelength), where variations are seen prominently in low-frequency observations \citep{donner+20, inpta22}. 
Spatial variations in the electron column density results in multipath propagation, resulting in delays scaling as $\lambda^\alpha$, with $\alpha \sim 4.0\pm0.6$ \citep{oswald+21}.
In the time domain, this effect can be seen through the broadening of pulses by a characteristic scattering tail (e.g. \citealt{bhat+04}, although with the additional complication that different scattered paths can encounter a different electron column \citep{cordes+16, donner+19}). 
In the Fourier domain this is observed as scintillation, where temporal and spectral variations of flux density arise from interference between deflected, coherent images of the pulsar. 

Pulsar scintillation is now commonly studied through the secondary spectrum, which is the 2D power spectrum of the scintillation pattern.
In this space, many sources have been seen to have `scintillation arcs' \citep{stinebring+01}, parabolic distributions of power which indicate scattering being dominated by highly localized regions, or `thin screens'  \citep{walker+04, cordes+06}.
Furthermore, the presence of sharp inverted parabolic `arclets' stemming from the main parabola are seen in some cases, implying strong anisotropy (seen in $\sim 20\%$ of sources in \citealt{stinebring+22}). 

The majority of the brightest known radio pulsars are isolated and slowly rotating \citep{psrcat05}, which due to high S/N requirements have been the focus of the widest-ranging studies of scintillation arcs to date 
\citep{stinebring+22, wu+22, main+23}.
However, recent studies have begun to show the power of studying scintillation arcs in millisecond pulsars. 
For the precision timing of PTA pulsars, scattering variations may be a source of uncorrected correlated noise \citep{goncharov+21b, chalumeau+22}, which can be estimated through scintillation arcs \citep{hemberger+08}, or through the frequency scale of scintillation (for applications to PTA data, see e.g. \citealt{levin+16, liu+22}). 
Additionally, scintillation arcs encode the relative velocity and distance of the pulsar, scattering screen, and the Earth, so modelling of their annual and orbital variations can be used to precisely measure pulsar orbital parameters as well as screen distances \citep{reardon+20, walker+22, mckee+22}. 

The Large European Pulsar Array (LEAP) is a 195-m tied-array beam telescope comprised of many of the largest telescopes in Europe, and has been observing $>20$ MSPs at monthly cadence since 2012 as part of the European Pulsar Timing Array (EPTA) \citep{stappers+kramer11, kramer+champion13, bassa+16}. Owing to its sensitivity and data products which can be re-reduced to any time and frequency resolution, LEAP is well-suited to study MSP scintillation.  In studies to date, secondary spectra have been used to measure the total time delays from scattering in PSR J0613$-$0200 (\citealt{main+20}), and to associate the scattering screen of PSR J1643$-$1224 with a known HII region \citep{mall+22}.

In this paper, we present the first large sample of scintillation arcs of MSPs, observed over the last 10 years with LEAP, and the Effelsberg 100-m telescope. The paper is organized as follows: in Section \ref{sec:background} we revisit the necessary theory of scintillation arcs for this work, in Section \ref{sec:data} we describe our observations and data, and in Section \ref{sec:analysis} we describe our analysis. We discuss the the results of particular pulsars in Section \ref{sec:results}, and the whole sample in Section \ref{sec:discussion}. Section \ref{sec:conclusions} contains the ramifications of our findings and prospects for the future.

\section{Background of scintillation}
\label{sec:background}

\subsection{Arc Curvature and Scintillation Velocity}

Here we briefly review the relevant theory of scintillation arcs, which we detailed in \citet{main+20} (originally developed, and explained in more detail in \citealt{walker+04, cordes+06}).  

The dynamic spectrum $I(t, \nu)$ is the measured flux density as a function of time and frequency (typically averaged over many pulses), showing variations owing to interstellar scintillation.
In the 2D power spectrum of the dynamic spectrum $S(f_{t}, f_{\nu}) = |\tilde{I}(f_{t}, f_{\nu})|^{2}$, referred to as the `secondary spectrum', the conjugate variable of time  $f_{t} \equiv f_{\rm D}$ is related to the Doppler shift between deflected paths, and depends on the angles $\theta$ of two deflected paths ($ij$) as 
\begin{equation}
f_{D, ij} = \frac{ (\boldsymbol{\theta}_{i} - \boldsymbol{\theta}_{j} ) \cdot \boldsymbol{v}_{\rm eff} }{\lambda},
\end{equation}
and the conjugate variable of frequency $f_{\nu} \equiv \tau$ is related to the delay between image pairs, described as 
\begin{equation}
\tau_{ij} = \frac{d_{\rm eff} (\theta_{i}^{2} - \theta_{j}^{2})}{2 c}.
\end{equation}
The effective distance $d_\mathrm{eff}$ and effective velocity $\boldsymbol{v}_\mathrm{eff}$ depend on the relative distances and velocities of the pulsar ($d_\mathrm{psr}$, $\boldsymbol{v}_\mathrm{psr}$), screen ($d_\mathrm{scr}$, $\boldsymbol{v}_{\rm scr}$), and Earth ($\boldsymbol{v}_{\oplus}$), as
\begin{align}
d_{\rm eff} &= (1/s - 1) d_{\rm psr}, \\
\boldsymbol{v}_{\rm eff} &= (1/s - 1) \boldsymbol{v}_{\rm psr} + \boldsymbol{v}_{\oplus} - \boldsymbol{v}_{\rm scr} / s,
\end{align}
where the fractional screen distance from the pulsar is defined as $s \equiv 1 - d_{\rm scr} / d_{\rm psr}$.

When one of the two scattered angles is 0 (i.e. the theoretical undeflected line of sight), the common dependence of $f_{\rm D}$ and $\tau$ on the observed angle $\theta$ results in a parabolic distribution of power 
\begin{equation}
\tau = \eta f_{\rm D}^2.
\end{equation}
The proportionality constant, or `arc-curvature' $\eta$ depends on the relative distances and velocities of the pulsar, screen, and Earth, as
\begin{equation}
\eta = d_{\rm eff} \lambda^{2} / 2c v_{\rm eff}^{2} \cos(\alpha)^{2},
\end{equation}
where $\alpha$ is the angle between $\boldsymbol{v}_{\rm eff}$ and $\psi$.

Throughout this paper, we work with the distance weighted effective velocity $W$, which rearranges $\eta$ to separate the unknown values, and is proportional to $|v_{\rm eff, \shortparallel}|$,
\begin{equation}
W \equiv \frac{|v_{\rm eff, \shortparallel}| }{ \sqrt{d_{\rm eff} } } = \frac{\lambda}{\sqrt{2c \eta}}.
\end{equation}
This is the same approach as in \citet{main+23}, and proportional to the quantity used in \citet{mall+22}; the main benefits of $W$ is that it does not diverge as $|v_{\rm eff, \shortparallel}| \rightarrow 0$ and is independent of the observing frequency. 

In the absence of arcs, a characteristic curvature $\eta$ can be estimated from the time and frequency scale from the 2D autocorrelation function (ACF) of scintillation.  The scintillation bandwidth $\nu_{\rm s}$ is defined as the half-width at half-maximum of the ACF in frequency, and the scintillation timescale $t_{\rm s}$ is defined as the $1/e$ point of the ACF in time \citep{cordes86}.

Using thin screen relations (with a phase structure function with index of 2, details in \citealt{cordes+98}), the corresponding time delay and Doppler shifts can be inferred from the scintillation bandwidth and timescale as
\begin{equation}
\tau_{\rm s} \approx 1/2\pi \nu_{\rm s}, \quad f_{\textrm{D}, \rm s} \approx 1 /\sqrt{2} \pi t_{\rm s}.
\end{equation}
These relations are approximate, as the prefactors are model dependent.  
Then we can estimate $W$ as
\begin{equation}
    W \approx \frac{\lambda}{t_{\rm s}}\sqrt{\frac{\nu_{\rm s} }{2c\pi}}.
\end{equation}
A measure of either the arc curvature $\eta$, or the scintillation timescale and bandwidth, are then a direct measure of $v_{\rm eff, \shortparallel} / \sqrt{d_{\rm eff}}$.  In this paper, we measure $W$ from the secondary spectrum wherever possible; scintillation velocities derived from $t_{\rm s}$ and $\nu_{\rm s}$ are dependent on the distribution of power across the screen, and will systematically vary as different regions of the same scattering screen are seen \citep{cordes+98, rickett+14, reardon+19}.  Measurements of scintillation arc curvatures are more robust, demonstrated to be stable to changes in screen's anisotropy \citep{reardon+20}, or in the presence of significant substructure \citep{sprenger+22}.

\subsection{Timescale for feature movement}
\label{sec:timescale}

Secondary spectra often show compact features at fixed angular positions, which over time are seen to travel through the secondary spectrum due to their relative velocity \citep{hill+05, sprenger+22}. 
An important quantity is the timescale for features to pass through the secondary spectrum, which sets a timescale for scattering delays to correlate. This timescale (effectively the same as the traditional ``refractive timescale'' of scintillation) of the screen is the time it takes to traverse to a new section of the screen.  For an observable portion of the screen with angle $\theta_0$ with corresponding size length $l_{\rm r}$,
\begin{equation}
    l_{\rm r} = 2 \theta_0 d_{\rm scr} = \sqrt{\frac{8c\tau_0}{d_{\rm eff}}} d_{\rm scr},
\label{eq:refractivescale}
\end{equation}
then the timescale is related to the arc curvature and the maximum delay $\tau_{0}$ as
\begin{equation}
    t_{\rm r} \approx \frac{l_{\rm r}}{s v_{\rm eff}} = \frac{\sqrt{8c\tau_0}}{W}. 
\end{equation}
In \citet{sprenger+22}, all of the observables were expressed as a function of the feature movement in $\sqrt{\tau}$, related to the distance weighted effective velocity as
\begin{equation}
\partial_t \sqrt{\tau} = \frac{W}{\sqrt{2c}} = \frac{1}{2\nu \sqrt{\eta}},
\label{eq:featuremovement}
\end{equation}
which can also be used to obtain equation \eqref{eq:refractivescale}.
The relevant $\tau_0$ value can either be the highest $\tau$ that features are visible, which gives the timescale for features to completely pass through the secondary spectrum, or the $1/e$ value of $\tau$. 

\section{Observations and Sample}
\label{sec:data}

\subsection{LEAP}
The Large European Array for Pulsars (LEAP) is a tied-array telescope comprised of the Effelsberg 100-m telescope,
the Lovell telescope at Jodrell Bank Observatory, the Westerbork Synthesis Radio Telescope, the Nan\c cay Radio Telescope and the Sardinia Radio Telescope, simultaneously observing MSPs with monthly cadence.  Each observing run has some subset of these telescopes; the voltage data are recorded at each site, then shipped to Jodrell Bank to be correlated and coherently added using the pipeline developed by the LEAP team (details of LEAP in \citealt{bassa+16}, and details of the correlator in \citealt{smits+17}).  When all telescopes are available, this results in an effective 195-m diameter dish. The scans are typically 30$-$60 minutes. The data are recorded in contiguous 16-MHz sub-bands covering 1332--1460\,MHz, where the total bandwidth per observation varies between 80--128\,MHz depending on the telescopes used; Jodrell Bank and Sardinia never use the full 128-MHz bandwdith.  While the standard folding pipeline produces archives with 10-s integrations, 1-MHz channels, the coherently added voltages are stored on tape, allowing us to later reduce the data with much higher spectral resolution to study fine scintillation features. 

\subsection{Effelsberg}

\subsubsection{Long Targeted Observations}

Several of the sources have scintillation which is barely resolved in a $\sim 30-60$ minute observation, and with scintles comparable to LEAP's 128-MHz band. This motivated us to take tailored observations with a wider bandwidth and a longer duration.  In addition to the LEAP observations, we obtained 2-3 hour observations of several LEAP sources where the scintillation was not quite resolved in frequency or time. The data were taken using the PSRIX backend (\citealt{lazarus+16}), using the central feed of the 7-beam receiver, recorded in 25\,MHz subbands with a usable bandwidth of 1250-1450\,MHz and saved in \textsc{psrdada} \footnote{\url{http://psrdada.sourceforge.net/}} format. 

\subsubsection{EPTA observations}
Since March 2021, Effelsberg observations for the EPTA record a separate parallel data stream suitable for scintillation studies, using the Effelsberg Direct Digitisation system, folding with 64000 channels across 1200--1600\,MHz (sensitive to delays of $\tau < 80\,\upmu$s).  These are now a regular data product in EPTA observations, and along with LEAP will double the cadence of observations suitable for studying scintillation arcs. 

\subsection{Sample}

In our sample, there are 6 sources with resolvable, variable arcs within the 30$-$60\,minute LEAP observations: PSRs J0613$-$0200, J0621+1002, J1600$-$3053, J1643$-$1224, J1918$-$0642, J1939+2134 (B1937+21). In these sources, we study the time variability of the arc curvature, and time delays.  A scintillation arc is also faintly seen in the highest S/N observation of PSR J1824$-$2452A (B1821$-$24A).

In addition, we also investigate sources where scintillation is marginally resolved by LEAP, supplementing this study with longer $2-3$\,hour Effelsberg observations. These include single scans on PSRs J0751+1807, J1713+0747, J1832$-$0836, J1857+0943 (B1855+09), J2010$-$1323, as well as a high-cadence approximately bi-weekly campaign on PSR J0613$-$0200 from March$-$June 2020, totalling 19 observations, which was previously included in \citet{main+20}.  The first year of EPTA fine-channel scintillation products are included for PSRs J0613$-$0200 and J1643$-$1224, with 12 and 9 observations respectively, to demonstrate the value of these data products to increase the cadence and timespan of scintillation data products.
A summary of the samples is given in Table \ref{tab:sample}.  
Our sample partially overlaps with the scintillation study of EPTA pulsars in \citet{liu+22}, where the larger bandwidth, but lower frequency resolution of the Nan\c cay dataset allowed for studies of slightly less scattered sources.  

\begin{table*}
 \caption{Summary of the pulsars in our sample.  The pulsar distances shown with $1\sigma$ errorbars are all timing parallax measurements compiled by the IPTA in \protect{\citet{perera+19}}, while the distances shown without errors are DM distance estimates from YMW16 model \protect{\citet{yao+17}}. The DM values are taken from psrcat \citep{psrcat05}.
 While LEAP has heightened sensitivity, the targeted Effelsberg observations have a larger bandwidth, and longer durations as described in Section \ref{sec:data}.  
 }
\begin{center}
\begin{tabular}{ lllllllllll } 
 \hline
 Pulsar Name & DM & P$_{\rm b}$ & $d_{\rm psr}$ & $\mu_{\alpha}$ & $\mu_{\delta}$ & $\langle$S/N$\rangle$ & $\langle t_{\rm obs} \rangle$ & $\nu_{\rm chan}$ & N$_{\rm phase}$ & Res.\\
   & (pc\,cm$^{-3}$) & (days) & (kpc) & (mas/year) & (mas/year) & & (min.) & (kHz) & & Scint\\
\hline
J0613$-$0200 & 38.78  & 1.20  & $1.11\pm0.05$ & $1.828(5)$ & $-10.35(1)$ & 134 & 51 & 62.5 & 128 & LEAP \\
J0621+1002   & 36.47  & 8.32  & $0.4\pm0.1$ & $3.2(1)$  & $0.6(5)$ & 71 & 39 & 62.5 & 128 & LEAP \\
J0751+1807   & 30.25  & 0.26  & $1.4^{+0.4}_{-0.3}$ & $-2.72(6)$ & $-13.4(3)$ & 140 & 180 & 100 & 128 & Eff \\
J1600$-$3053 & 52.33  & 14.35 & $2.0^{+0.3}_{-0.2}$ & $-0.97(1)$ & $-7.04(5)$ & 139 & 54 & 31.125 & 64 & LEAP \\
J1643$-$1224 & 62.41  & 147.02& $1.1^{+0.6}_{-0.3}$ & $6.03(3)$ & $4.1(1)$ & 231 & 33 & 7.8125 & 16 & LEAP \\
J1713+0747   & 15.92  & 67.83 & $1.20\pm0.03$ & $4.924(1)$ & $-3.913(2)$ & 408 & 180 & 100 & 128 & Eff \\
B1821$-$24A  & 119.89 & -     & $3.7\pm0.7$ & $-0.2(2)$ & $-6(4)$ & 132 & 44 & 31.125 & 64 & LEAP \\
J1832$-$0836 & 28.19  & -     & $0.8\pm0.2$ &  $-7.97(5)$ & $-21.2(2)$ & 55 & 120 & 100 & 128 & Eff \\
B1855+09     & 13.31  & 12.33 & $1.1\pm0.1$ & $-2.652(4)$ & $-5.423(6)$ & 138 & 180 & 100 & 128 & Eff \\
J1918$-$0642 & 26.46  & 10.91 & $1.3^{+0.2}_{-0.1}$ & $-7.15(2)$ & $-5.94(5)$ & 53 & 30 & 125 & 128 & LEAP  \\
B1937+21     & 71.02  & -     & $4.7^{+1.4}_{-0.9}$ & $0.074(2)$ & $-0.410(3)$ & 485 & 45 & 31.125 & 32 &  LEAP \\ 
J2010$-$1323 & 22.18  & -     & $1.9^{+0.8}_{-0.5}$ & $2.56(6)$ & $-5.9(2)$ & 82 & 180 & 100 & 128 & Eff\\
\hline
\end{tabular}
 \label{tab:sample}
\end{center}
\end{table*}

\section{Scintillation data products and measurements}
\label{sec:analysis}

\subsection{Dynamic and secondary spectra}
The creation of dynamic and secondary spectra are almost identical to the methods described in \citet{main+20, mall+22}, however we briefly review and describe differences here.  

Data were folded using the \texttt{dspsr} software package \citep{vanStraten+11}, beginning from the baseband data. 
The time, phase, and frequency bins were different for each specific pulsar, chosen to fully resolve the scintillation in time and frequency, or equivalently, to extend sufficiently far in $f_{\rm D}$ and $\tau$ to capture the full extent of arcs in the secondary spectrum.
The sub-bands were combined in frequency using \texttt{psradd} in the \texttt{psrchive} software package \citep{hotan+04}, and polarisations were summed to form total intensity. Radio Frequency Interference (RFI) flagging and treatment of masked pixels is identical to \citet{main+20}.  The outer $10\%$ of the dynamic spectra are tapered with a Hanning window, before forming the secondary spectrum.

As the Effelsberg observations cover a larger bandwidth, we use the `NuT transform' instead of a direct FFT in time to form secondary spectra, as described in \citet{sprenger+21}.  This transformation is a direct Fourier Transform over a scaled time axis of $t^{'} = t \nu / \nu_{\rm ref}$ in every channel with frequency $\nu$. This prevents smearing of scintillation arcs owing to the $\eta \propto \lambda^2$ dependence, and ensures the contribution to scintillation from sources of fixed angular position are at fixed position in the secondary spectrum of a single observation. We apply this in all of our observations, referencing to $\nu_{\rm ref}=1400\,$MHz.  We note that this transform can lead to some artefacts in the secondary spectrum, as power on the $\tau=0$ axis (arising from e.g. RFI, pulse-to-pulse variations) is spread to diagonal lines. We do not see these artefacts prominently in our observations, as they are most prominent in observations with long-durations, large fractional bandwidths, and large pulse-to-pulse flux variations.

\begin{figure*}
\centering
\includegraphics[width=0.9\textwidth]{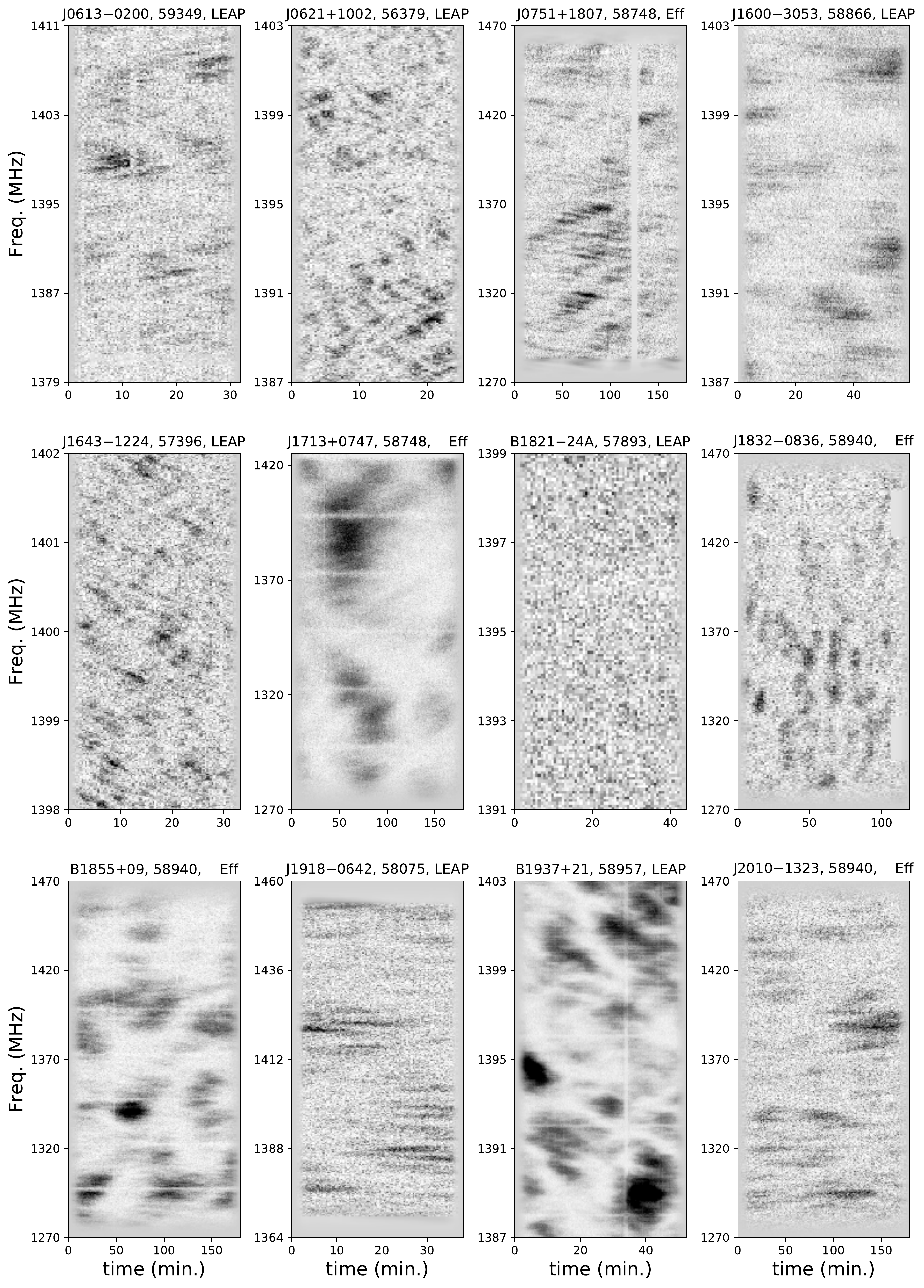}
\caption{ Dynamic spectra of pulsars in the sample.  Several of the LEAP sources, only a subset of the band is shown to display the frequency scale of the scintles. The mean of each dynamic spectrum is normalized to 1, and the intensity map covers the range -1$-$6.}
\label{fig:Dynspecs}
\end{figure*}

Representative dynamic spectra of the sources in our sample are shown in Figure \ref{fig:Dynspecs}, and their associated secondary spectra are shown in Figure \ref{fig:LEAPpano}. The secondary spectra of every observation used in this work are included in the Appendix.

\begin{figure*}
\centering
\includegraphics[width=0.99\textwidth]{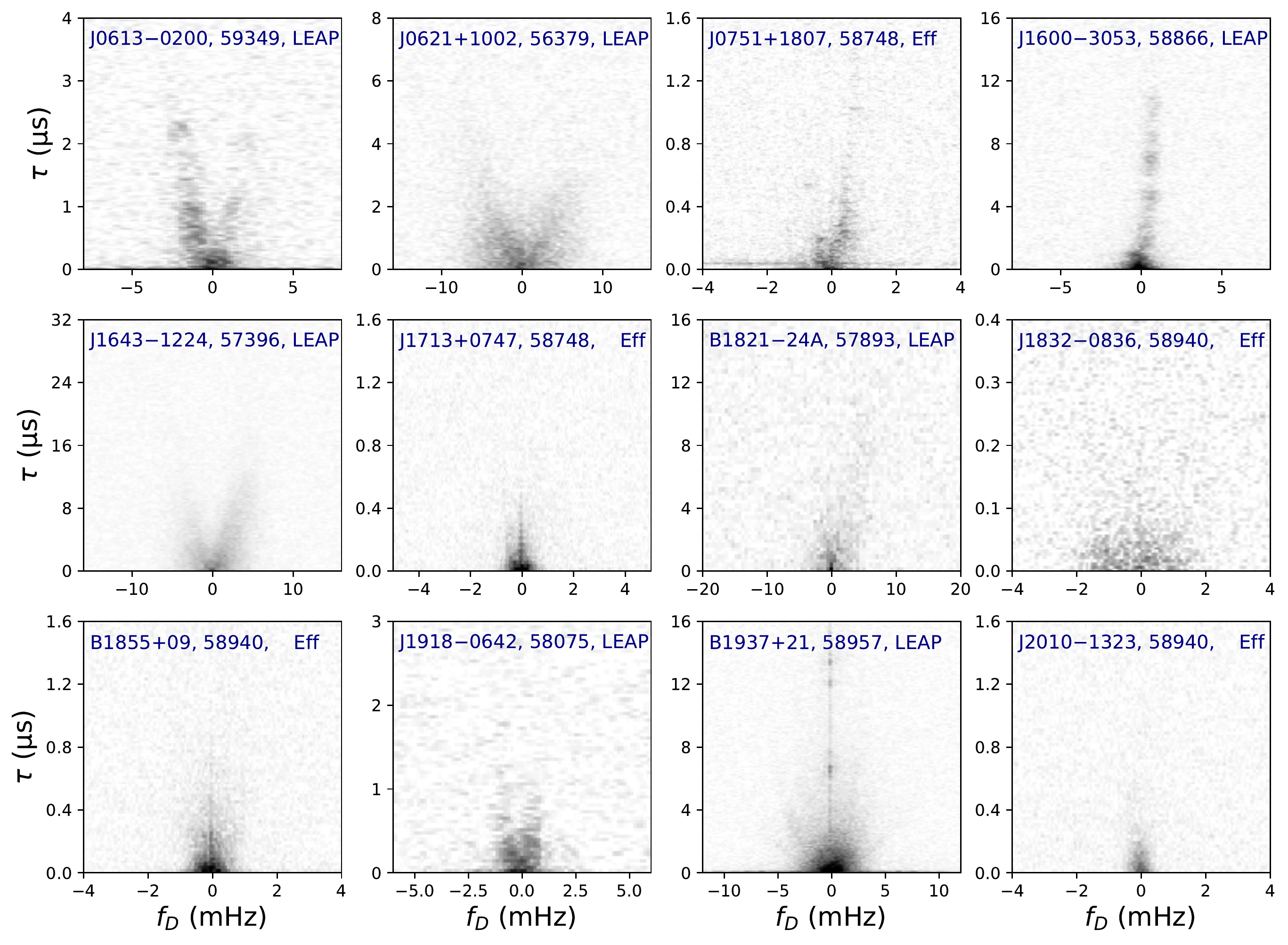} \\
\vspace{-2mm}
\caption{ Panorama of scintillation arcs, from the corresponding dynamic spectra in Figure \ref{fig:Dynspecs}.  Scintillation arcs are seen in many sources, with highly varied extents in $\tau$ and $f_{\rm D}$, owing to differences in the scattering screens, and relative velocities. The arcs are qualitatively different between sources, showing compact regions of power in some (e.g. PSRs J0613$-$0200, J1600$-$3053, J0751+1807), clear parabolae in others (e.g. PSRs J0621+1002, J1643$-$1224), and multiple parabolae in PSR B1937+21.  The results are described in more detail in section \ref{sec:results}. The intensity maps are logarithmic, extending 3 orders of magnitude in most cases, or 1 in B1821$-$24A }
\label{fig:LEAPpano}
\end{figure*}

\subsection{Arc curvature Measurement}

We have measured the arc curvature using the ``normalized secondary spectrum'' as done in \citet{reardon+20, walker+22}, in which the $f_{\rm D}$ axis of the secondary spectrum is mapped to $f_{\rm D, norm} = f_{\rm D} \sqrt{\tau_{\rm ref} / \tau}$ (where we set the arbitrary reference time delay $\tau_{\rm ref} = \tau_{\rm max}$ throughout). This transformation effectively stretches the $f_{\rm D}$ axis, mapping parabolae to vertical lines of constant $f_{\rm D, norm}$.  Then $W$ can be identified by finding peaks in $S(f_{\rm D, norm})$ after performing a weighted sum over $\tau$. 

\begin{figure}
\centering
\includegraphics[width=1.0\columnwidth, trim=11cm 0cm 10.5cm 0.8cm, clip=true]{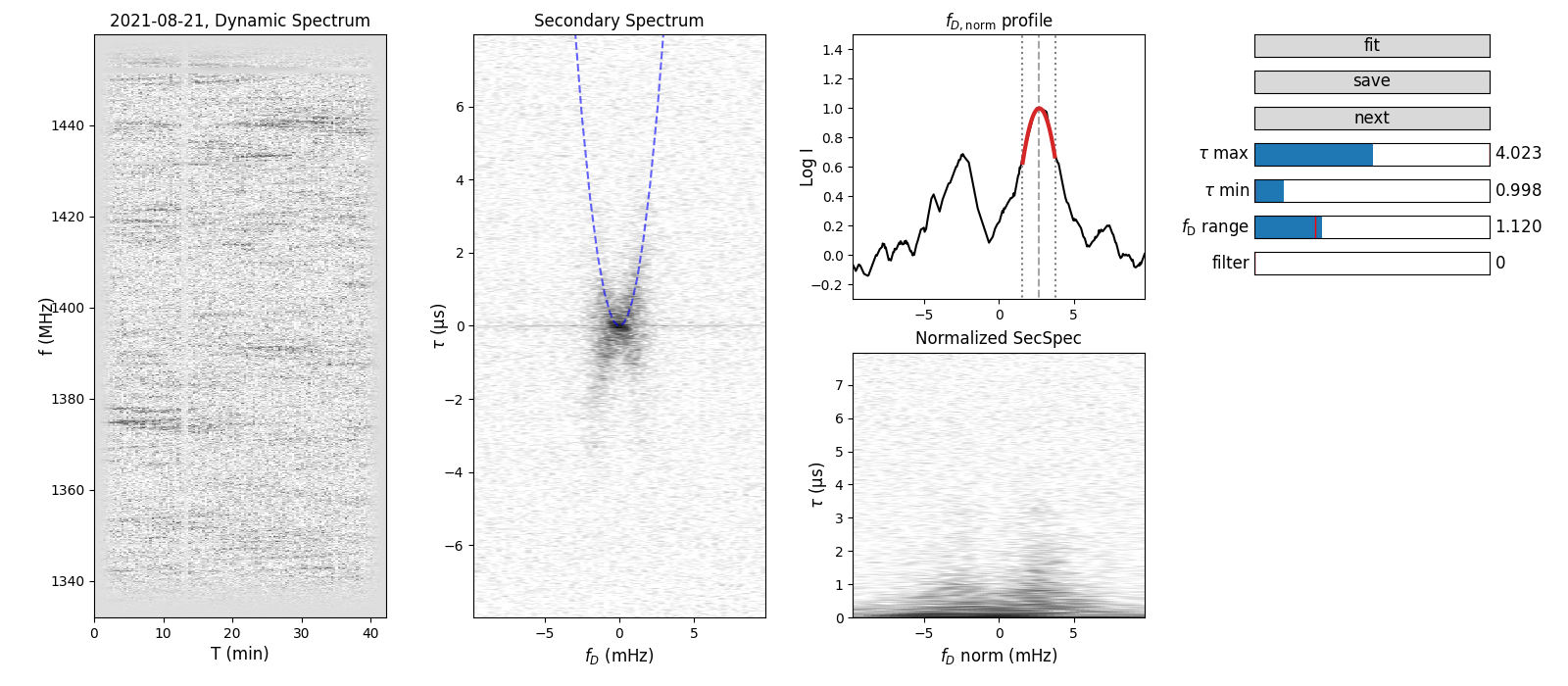}
\vspace{-0.6cm}
\caption{ An example of the arc curvature fitting algorithm from a LEAP observation of PSR J0613$-$0200. \textit{Left:} secondary spectrum, with best-fit parabola overlaid as the blue dashed line. \textit{Bottom-right:} normalized secondary spectrum. \textit{Top-right:} weighted sum of the normalized secondary spectrum over $\tau$.  The center and range of $f_{\rm D, norm}$ being fit is shown by the dashed and dotted grey lines, respectively, and the red parabola shows the fit. }
\label{fig:interactive}
\end{figure}

Arcs blend together at low values of $\tau$, becoming more clearly demarcated at high values of $\tau$.   
As the optimal range of $\tau$ to sum over and $f_{\rm D, norm}$ to fit vary between pulsars and between epochs, we applied the arc curvature fitting algorithm interactively.  Alongside the fitting, the dynamic and secondary spectrum of each observation are verified for corruption by RFI or phasing artefacts.  The range in $\tau$ to sum to form $S(f_{\rm D, norm})$ is adjusted per source, and peaks in $f_{\rm D, norm}$ can be given as initial guesses for a least-squares fit of a parabola to the local region of $S(f_{\rm D, norm})$, and the arc curvature is given as $\eta = f_{\rm D, norm}^2 / \tau_{\rm ref}$.  An example is shown in Figure \ref{fig:interactive}.

\subsection{Time delays from secondary spectra}

\begin{figure*}
\centering
\includegraphics[width=1.0\textwidth, trim=1.2cm 9.6cm 1cm 1cm, clip=true]{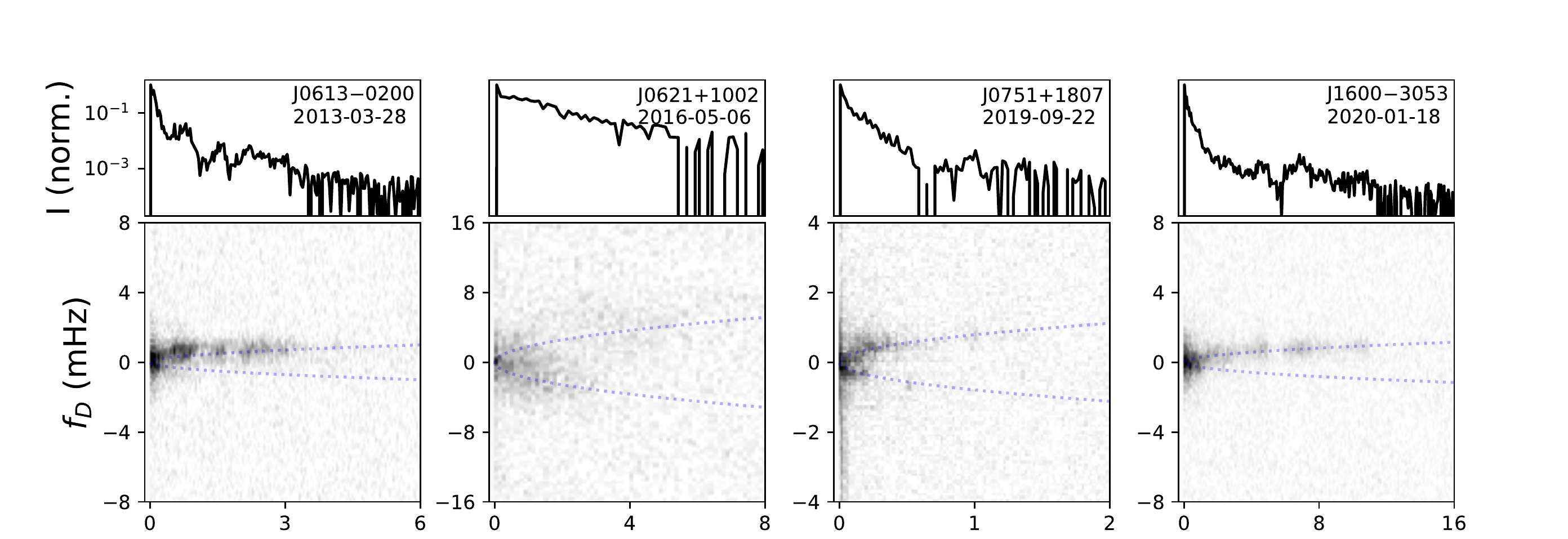} \\
\includegraphics[width=1.0\textwidth, trim=1.2cm 1cm 1cm 9.6cm, clip=true]{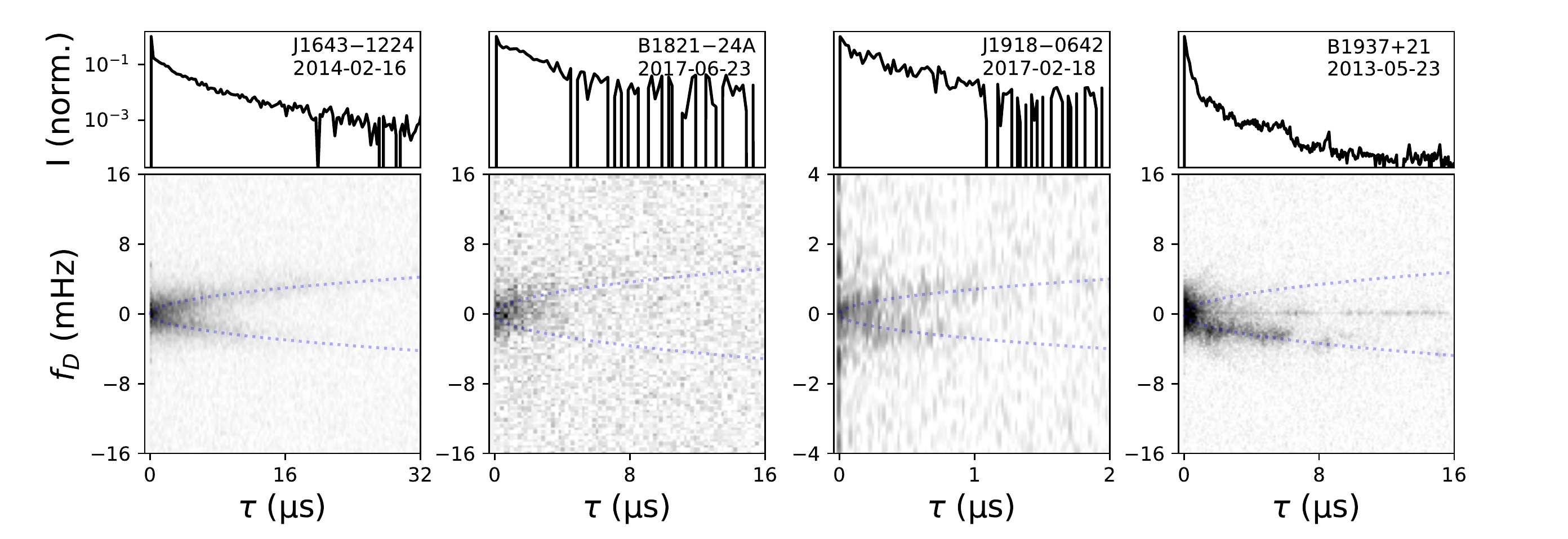} \\
\vspace{-10pt}
\caption{ Secondary Spectra (images), and the associated estimates of the scattering tail $I(\tau)$ by summing over $f_{\rm D}$ along the arc (top panels).  Several of the sources show features at large time delays, beyond a simple exponential tail.}
\label{fig:SecspecTails}
\end{figure*}

Secondary spectra express the power of scintillation in terms of the conjugate variables of time and frequency respectively, $f_{\rm D}$ and $\tau$.  In the strong scattering regime, the secondary spectrum contains contributions from all pairs of interfering images. \citet{hemberger+08} showed how one can estimate the averaged time delay $\langle \tau \rangle$ from the secondary spectrum, and in \citet{main+20} it was argued that this technique is valid in the limit of a strong, central image arising from a single thin screen, or when the response function is close to an exponential. The total geometric time delay $\langle \tau \rangle$ can then be estimated as
\begin{equation}
\langle \tau \rangle = \frac{ \int_{0}^{T} \tau |I(\tau)|^{2} d\tau}{\int_{0}^{T} |I(\tau)|^{2} d\tau}.
\label{eq:tauint}
\end{equation}
For each source, a range in $f_{\rm D}$ was chosen to fully encompass the power of the visible scintillation arc, and the background noise was estimated from a region of the same size offset by $f_{\rm D} \pm 30$\,mHz. The integrated profile of $\langle \tau \rangle$ against maximum $\tau_{\rm max} \equiv T$ was examined in all cases; the value and the error on $\langle \tau \rangle$ were estimated by the mean and standard deviation of the profile once it plateaus, taken in the range of $3T/4 < \tau < T$. Examples of secondary spectra and their associated profiles of $I(\tau)$ are shown in Figure \ref{fig:SecspecTails}.

\subsection{Scintillation Parameters from the ACF}

\begin{figure}
\centering
\includegraphics[width=0.526\columnwidth, trim=0cm 1.4cm 1cm 1cm, clip=true]{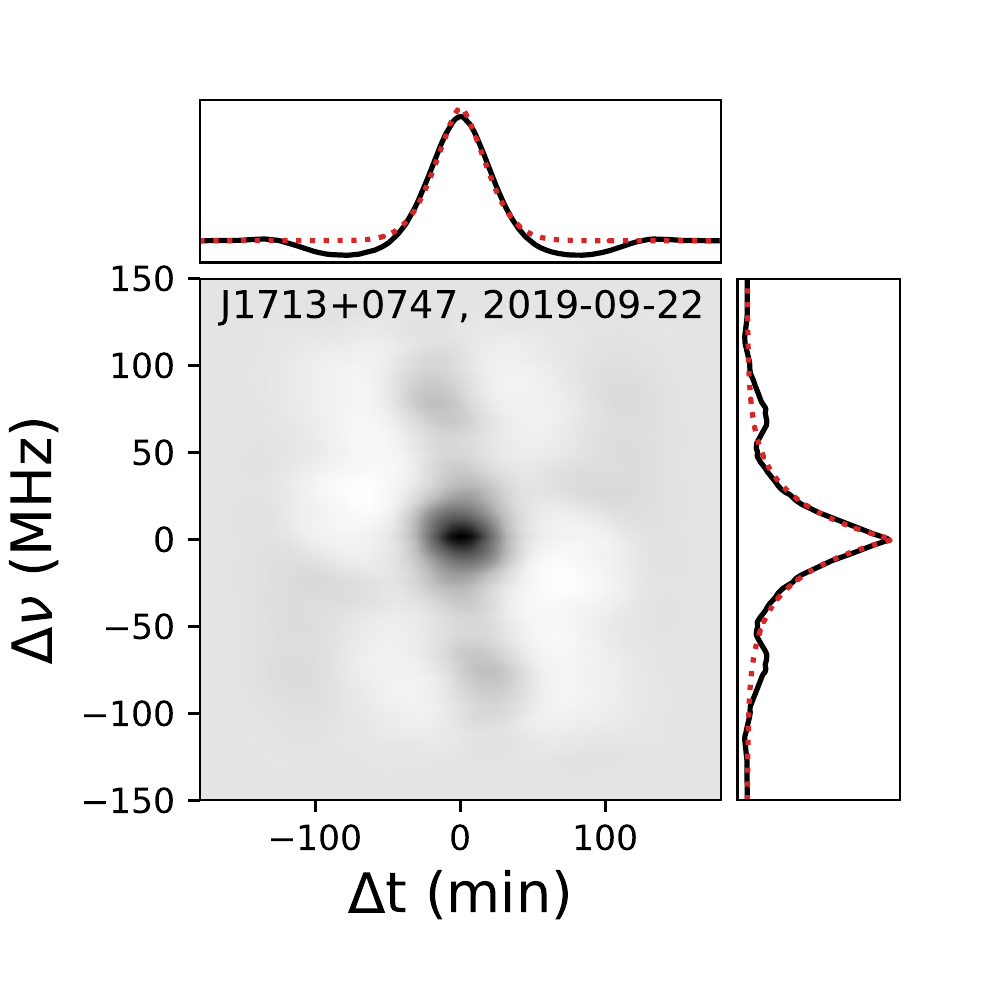}
\includegraphics[width=0.458\columnwidth, trim=1.2cm 1.4cm 1cm 1cm, clip=true]{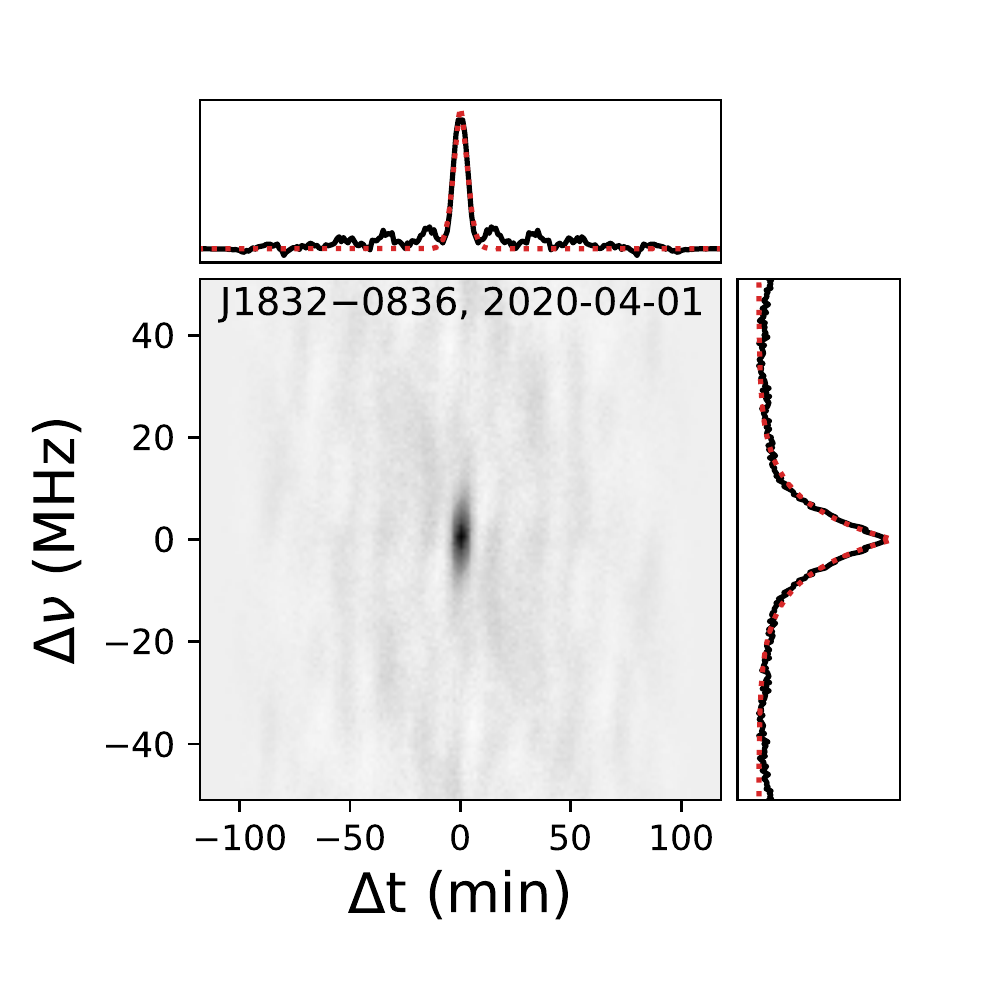} \\
\includegraphics[width=0.526\columnwidth, trim=0cm 0cm 1cm 1cm, clip=true]{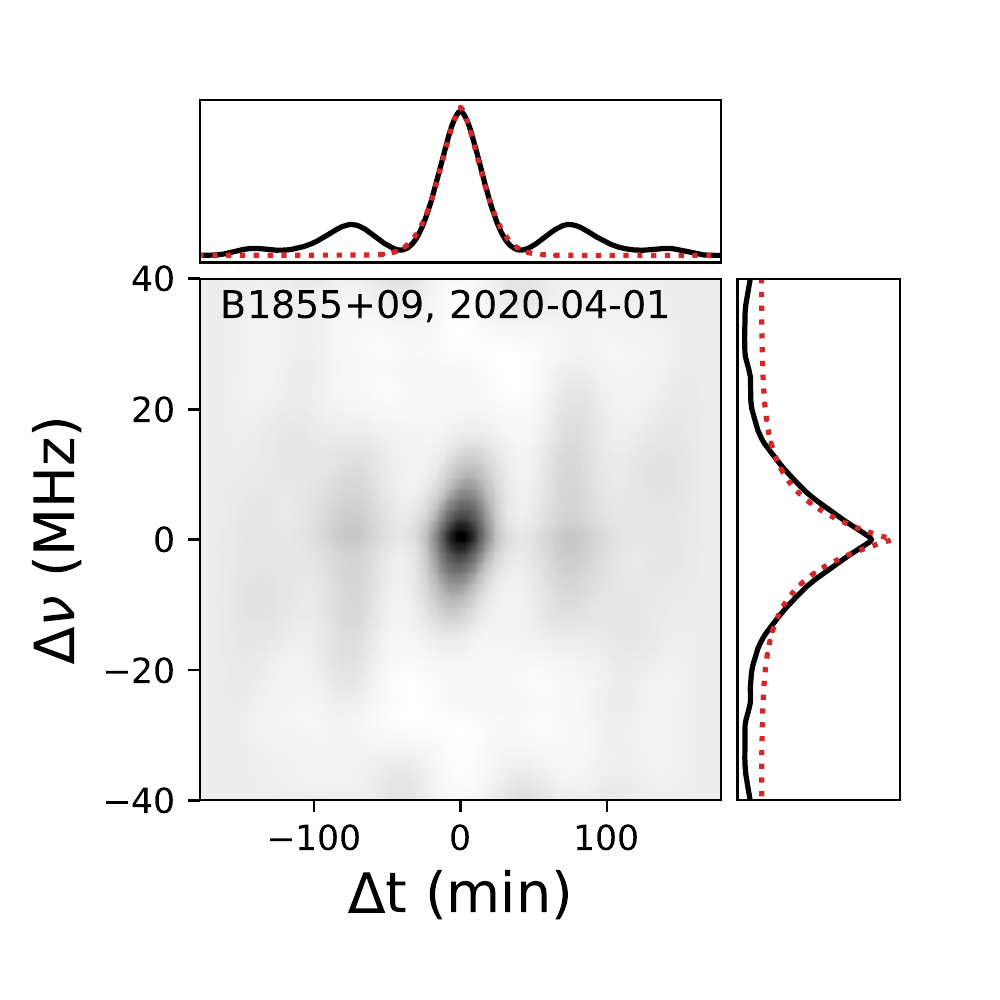}
\includegraphics[width=0.458\columnwidth, trim=1.2cm 0cm 1cm 1cm, clip=true]{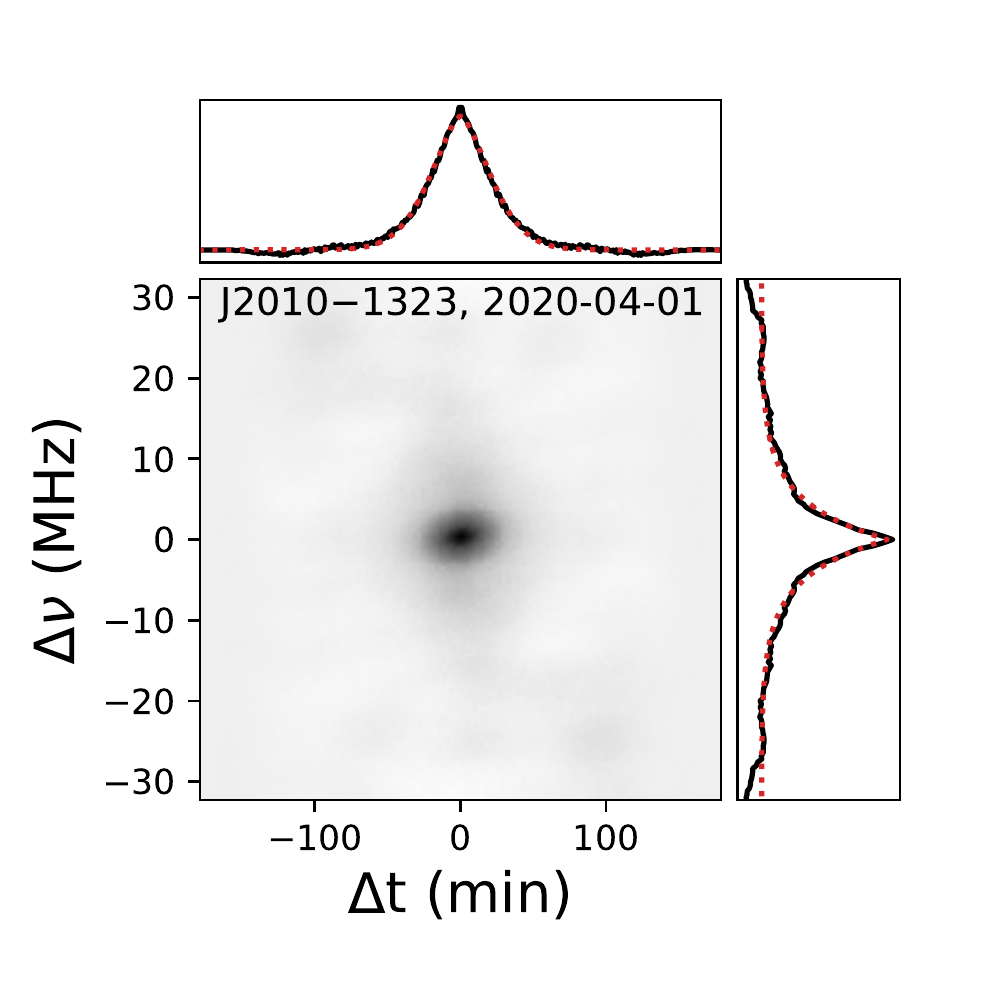}
\caption{ \textit{Images:} 2D ACFs $R(\Delta \nu, \Delta t)$ of the 4 pulsars without clear arcs.  \textit{Side panels:} cuts of $R(\Delta \nu, 0)$, $R(0, \Delta t)$ through the 2D ACF (solid black line), and best fit models from which the scintillation bandwidth and timescale are derived (dotted red line). }
\label{fig:ACFpano}
\end{figure}

In some of the sources with diffuse arcs, it is difficult to determine parameters from the secondary spectra.  In these cases, we measure the scintillation timescale and bandwidth in a more traditional way, through the 2D autocorrelation function (ACF) of the dynamic spectrum $R(\Delta \nu) = (I * I) (\Delta \nu)$. 
The ACF in frequency is fit with a Lorentzian, and with a Gaussian in time \citep{cordes86}.  The scintillation bandwidth is inversely proportional to the bulk scattering delay as $\langle\tau \rangle = C / 2\pi\nu_{\rm s}$, where the model-dependent constant $C$, depending on the distribution of scattered power, is assumed to be 1, which is the value for a thin screen with a square-law phase structure function \citep{cordes+98}.  The 2D ACFs of sources without clear arcs is shown in Figure \ref{fig:ACFpano}.

The precision on the measurements on the ACF is high, and the true error is dominated by the fact that there are finite scintles within the observation.  For an observation of duration $T_{\rm obs}$ with bandwidth BW, the fractional error is given by
\begin{equation}
    \sigma_s \approx \left(f_d  \frac{T_{\rm obs}}{t_{\rm s}} \frac{\rm{BW}}{\nu_{\rm s}}\right)^{-1/2},
\end{equation}
where the filling factor is assumed to be $f_d \approx 0.2$ \citep{cordes86}.

\section{Results for Individual Sources}
\label{sec:results}

\begin{figure*}
\centering
\includegraphics[width=0.44\textwidth, trim=0cm 0.8cm 0cm 0.0cm, clip=true]{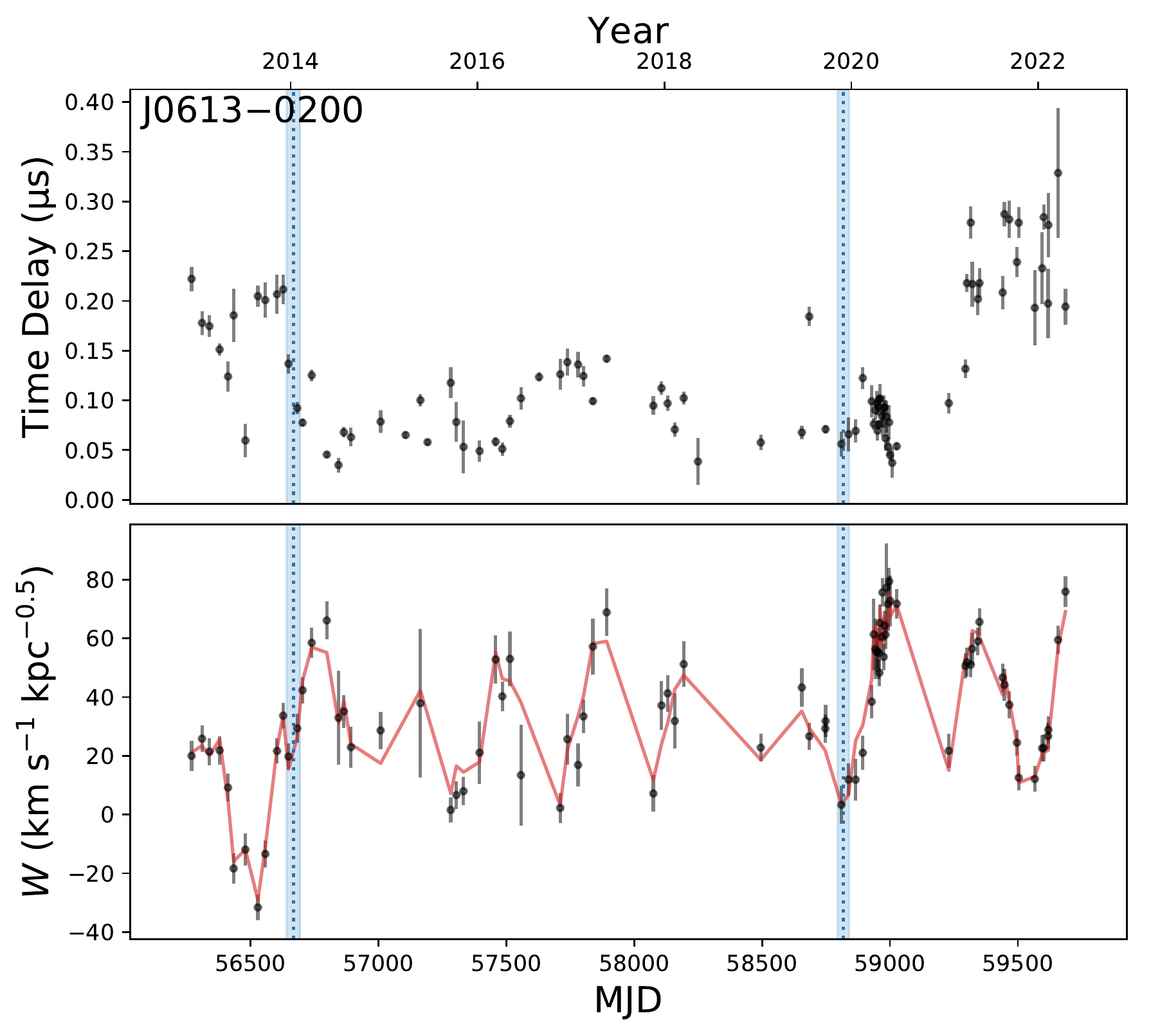} 
\hspace{5mm}
\includegraphics[width=0.44\textwidth, trim=0cm 0.8cm 0cm 0.0cm, clip=true]{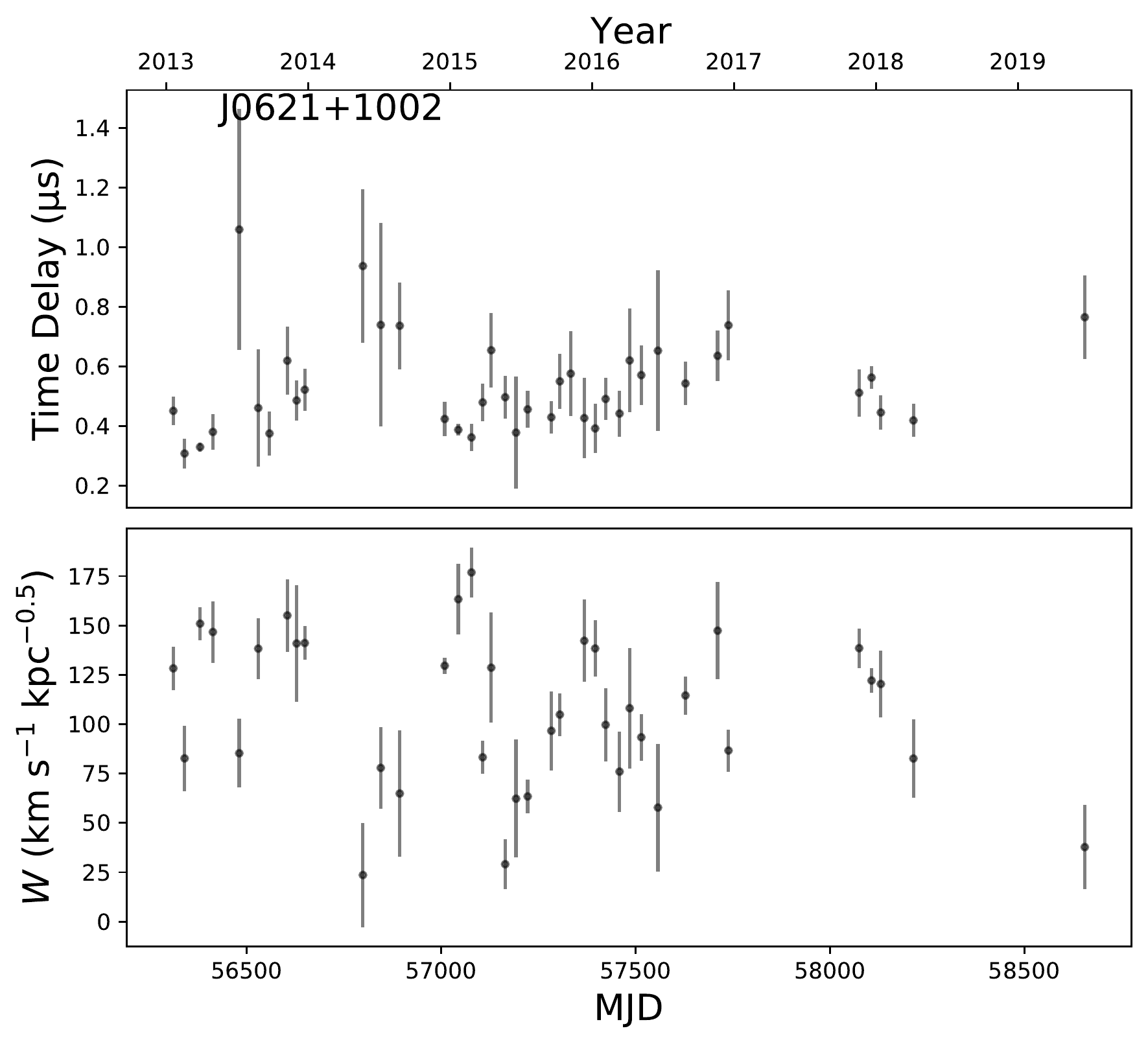} \\
\vspace{2mm}
\includegraphics[width=0.44\textwidth, trim=0cm 0.8cm 0cm 0.75cm, clip=true]{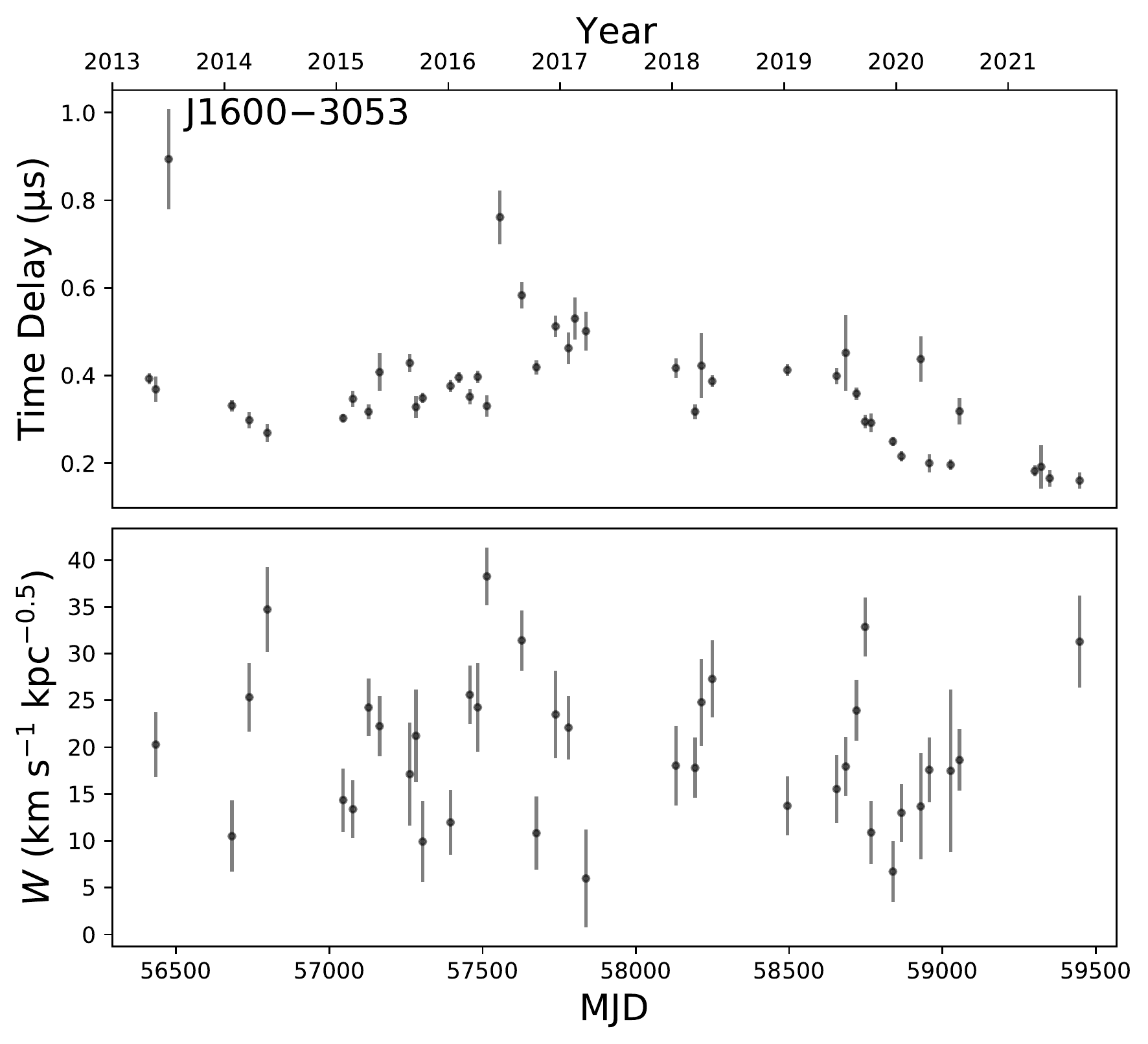} 
\hspace{5mm}
\includegraphics[width=0.44\textwidth, trim=0cm 0.8cm 0cm 0.75cm, clip=true]{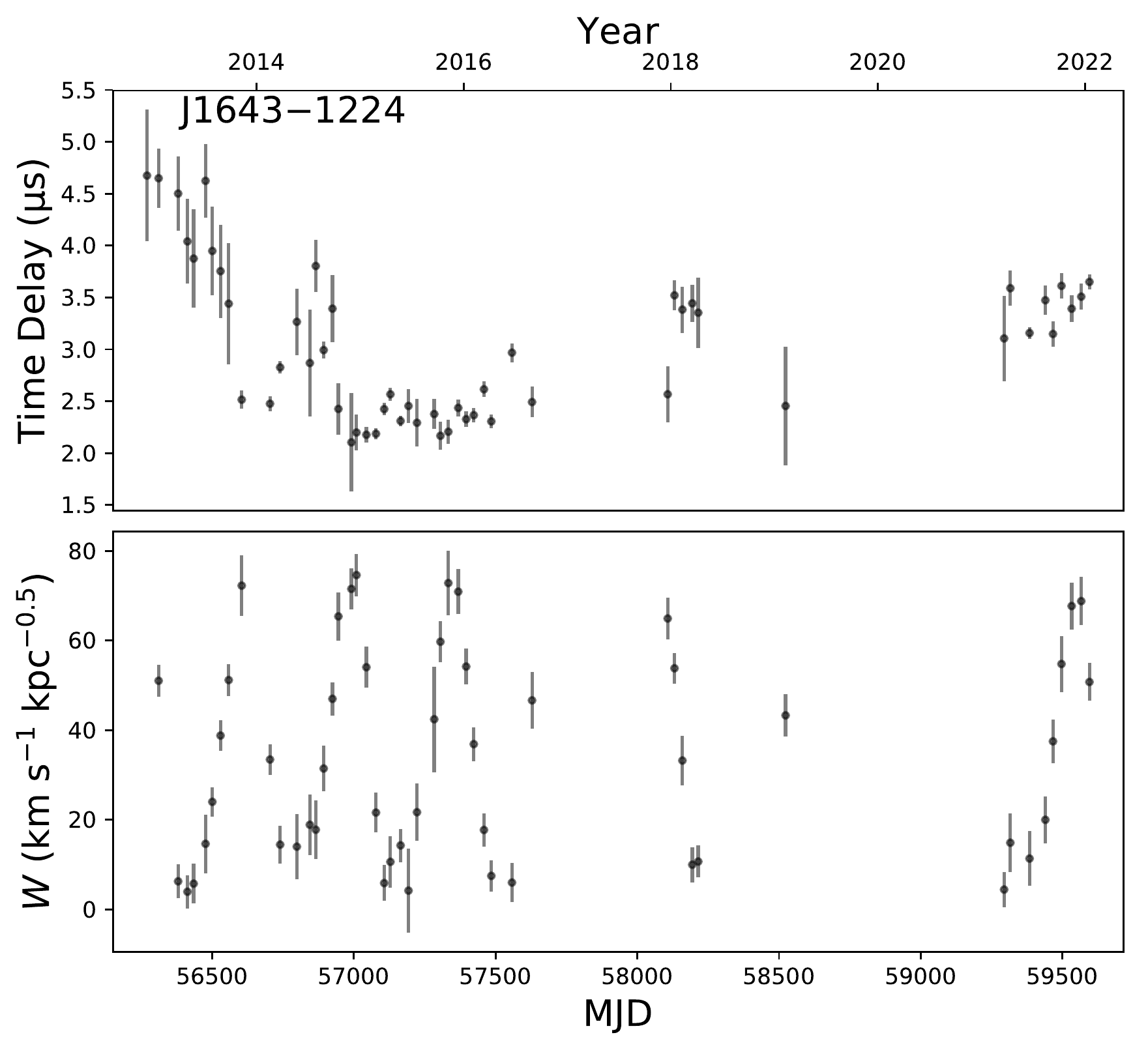} \\
\vspace{2mm}
\includegraphics[width=0.44\textwidth, trim=0cm 0.0cm 0cm 0.75cm, clip=true]{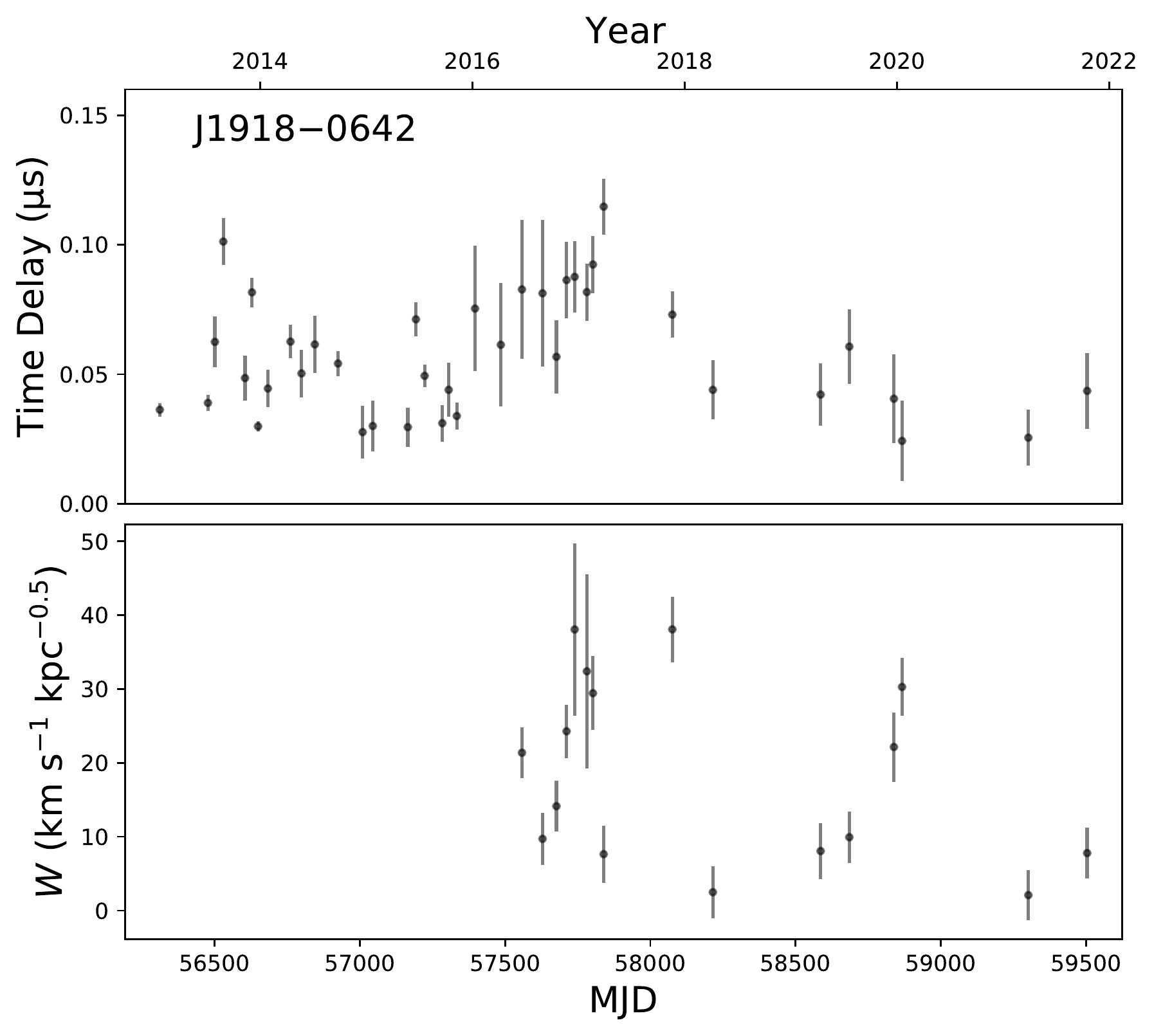} 
\hspace{5mm}
\includegraphics[width=0.44\textwidth, trim=0cm 0.0cm 0cm 0.75cm, clip=true]{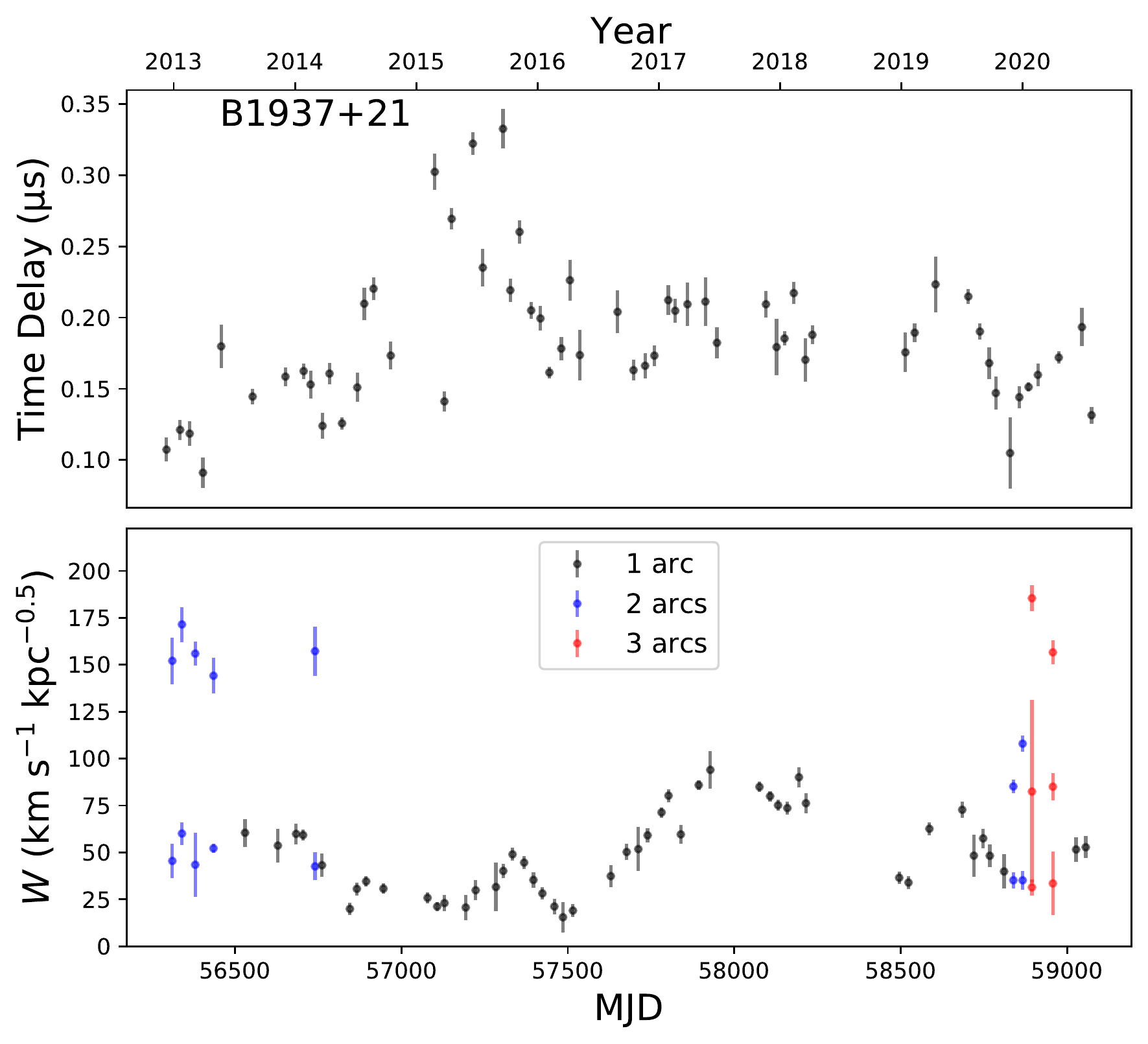} \\
\caption{ Time delays (\textit{Top}), and measurements of $W$ (\textit{Bottom}) for the 6 sources with fully resolvable scintillation at LEAP. For PSR J0613$-$0200, the fit of the annual and orbital variations of $W$ is shown in red, while the jumps of $\psi$ are shown in blue (fitting described in section \protect{\ref{sec:annualorbital}}). For PSR B1937+21, the different colours denote when 1, 2, or 3 parabolae can be identified.  }
\label{fig:timedelays}
\end{figure*}

The variable time delays and arc curvatures of all of our sources are shown in Figure \ref{fig:timedelays}, and a compilation of derived results are in Table \ref{tab:measurements}.  In this section we describe the results for specific sources.

\begin{table}
\begin{center}
\caption{Summary of quantities derived from the arc curvatures, and from ACF fits. The values of $W \equiv v_{\rm eff, \shortparallel} / \sqrt{d_{\rm eff}}$ and $t_{\rm r}$ quoted here are an average, representative value, and both will be time-variable owing to the annual motion, pulsar binary motion, and changing screen properties.  The values of $t_r$ are computed as in Section \ref{sec:background}, and represent the time for a fixed feature to pass from $-\langle \tau \rangle$ to $\langle \tau \rangle$ given the average value of $W$. }
\label{tab:measurements}
\begin{tabular}{ llll } 
\hline
 Pulsar Name & $\langle \tau \rangle$ & $\langle W \rangle$ & $\langle t_{r} \rangle$ \\
    & ($\upmu$s) & (km s$^{-1}$ kpc$^{-0.5}$) & (days) \\
\hline
J0613$-$0200 & $0.13 \pm 0.02$ & $34 \pm 3$ & $33 \pm 3$  \\
J0621+1002 & $0.47 \pm 0.02$   & $111 \pm 6$ & $19\pm1$ \\
J0751+1807 & $0.62\pm0.04$ & $35\pm2$   & $70\pm5$    \\
J1600$-$3053 & $0.34 \pm 0.02$ & $20 \pm 1$ & $93 \pm 7$ \\
J1643$-$1224 & $2.8 \pm 0.1$   & $33 \pm 3$ & $157 \pm 15$ \\
J1713+0747 & $0.010\pm0.003$ & $74\pm21$ &  $4.2\pm1.3$    \\
B1821$-$24A &  $1.0 \pm 0.5$  & $36 \pm 16$ & $88\pm45$ \\
J1832$-$0836 & $0.031\pm0.002$ & $240\pm16$ & $2.3\pm0.2$ \\
B1855+09 & $0.040\pm0.004$ & $45\pm5$ & $14\pm2$       \\
J1918$-$0642 & $0.05 \pm 0.01$ & $16 \pm 3$ & $45\pm9$ \\
B1937+21 & $0.18 \pm 0.01$ & $57 \pm 5$ & $23\pm2$ \\ 
J2010$-$1323 & $0.049\pm0.006$ & $28\pm3$ & $25\pm3$   \\
\hline
\end{tabular}
\end{center}
\end{table}

\subsection{Isolated Pulsars}

There are two isolated pulsars in the sample showing scintillation arcs, PSR B1937+21, and PSR B1821$-$24A.  They are in principle useful control sources, where for a fixed screen, variations in $W$ should arise only from Earth's motion.  Despite this, they show a range of interesting behaviour owing to dynamic screens.

\subsubsection{PSR B1937+21}

In low frequency observations of PSR B1937+21, there is previous evidence of a broad scintillation arc \citep{walker+13}. 
In our observations, we detect between one and three screens in a given observation.  The secondary spectra are largely devoid of structure in most observations, possibly due to convolution of multiple interacting screens, although an exception is shown in Figure \ref{fig:SecspecTails}, with distinct structures along the primary arc at delays $>8\upmu$s.  During certain ranges of time, there does appear to be a dominant screen showing annual variation.  However, the phase of the annual curve is not consistent over time, suggesting that the screen orientation is not fixed, or that different screens are varying at different times, shown in Figure \ref{fig:timedelays}. The curvature of the secondary and tertiary screens appear roughly consistent whenever they do reappear, suggesting stability over the 8 years of observations.  However, due to the difficulty of unambiguously identifying scattering screens, as well as the large distance uncertainty of the source, annual fits were not performed.

PSR B1937+21 often emits intrinsically narrow and bright ``giant pulses'', used in previous LEAP data to directly measure the time-variable scattering timescale \citep{mckee+19}.  Measurements of giant pulse scattering are direct, and unaffected by multi-screen effects.  Reassuringly, the trends of $\langle \tau \rangle$ over time is broadly in agreement between what was observed with giant pulses and the values derived from scintillation arcs in this work, with scattering rising from $\sim 0.2\,\upmu$s to $\sim 0.5\,\upmu$s in late 2014.  This serves as a useful cross-check of both methods.  The scattering seems to significantly vary from observation to observation implying variations on less than a month; variations in the DM of PSR B1937+21 have been observed on this timescale through high-cadence DM measurements \citep{chimepulsar21}.

\subsubsection{PSR B1821$-$24A}

PSR B1821$-$24A is a millisecond pulsar in the globular cluster M28.  The source emits giant pulses, and has been seen to have variable scattering times at L-band, with values as large as $25\pm8\upmu$s \citep{bilous+15}.  While this source has a low average signal-to-noise, we were able to detect a faint arc in our highest S/N observation, with time delays extending to $\sim 12\,\upmu$s.  Similarly to PSR B1937+21, this source could be a useful control for measuring time delays through scintillation or directly using giant pulse scattering.

The scattering appears to be consistent with the Milky Way ISM, rather than the intra-cluster gas.  Scattering in the intracluster gas would result in very large values of $W$; approximating $v_{\rm pl}$ with the core radius of $R_{c} \approx 0.37\,$pc, and $v_{\rm pl}$ with the velocity dispersion of $\approx 11\,$km\,s$^{-1}$ \citep{oliveira+22, baumgardt+18}, would result in $W \approx 600$\,km\,s\,kpc$^{-0.5}$, much greater than our measured $W = 36\pm16$\,km\,s\,kpc$^{-0.5}$. Comparably, our measured value of $W$ is easily compatible with typical Earth and screen velocities, at a screen of $\sim 1\,$kpc.  Additionally, NE2001 predicts of $\tau = 2.1\,\upmu$s \citep{cordes+02}, comparable to our measured $\langle \tau \rangle = 1.0\pm0.5\,\upmu$s.  

\subsection{Binary MSPs}

\subsubsection{PSR J0613$-$0200}

The scintillation of this pulsar was previously studied in \citet{main+20}, using LEAP and Effelsberg data from 2013$-$2020.  The time delays in 2013 were much larger than in subsequent years, corresponding also to a different screen orientation.  In the last two years of observations from 2020$-$2022, including EPTA scintillation data in the past year, the source has experienced heightened scattering fluctuations, with $\langle\tau\rangle$ decreasing to its lowest state in late 2020, and rising to its highest yet state in 2021 and beyond.  This change in scattering properties likely corresponds a change in the observed scattering screen, where either a different screen becomes dominant, or the screen's properties are changing. The modelling of the variable arc curvature of this source will be covered in Section \ref{sec:annualorbital}.  The secondary spectra often show distinct, compact features of $\lesssim 0.2 \upmu$s in extent along the main parabola, which can be tracked between observations during our high-cadence Effelsberg campaign in March$-$June 2020, covered in section \ref{sec:featuremovement}.

\subsubsection{PSR J0621+1002}

This pulsar shows well resolved, low curvature (i.e. large $W$) scintillation arcs, with faint indications of inverted arclets suggesting an anisotropic screen \citep{walker+04,cordes+06}. The time delays are of order $\sim 0.5\,\upmu$s, but with large measurement uncertainties due to low S/N per pixel in the secondary spectra arising from the diffuse arcs. The arcs are often featureless but highly anisotropic, changing on the timescale of months.  This likely reflects large, time-variable DM gradients across the screen, discussed in Section \ref{sec:phasegrad}.

This source is an ideal target for annual and orbital fitting of arc curvature (e.g. \citealt{reardon+20, mall+22}, see Sec. \ref{sec:annualorbital}). The advance of periastron $\dot{\omega}$ is significantly detected in timing \citep{perera+19}, which will allow for the component masses to be disentangled when combined with an inclination measurement from scintillation. Additionally, the well-resolved arcs also may enable novel techniques such as the $\theta$-$\theta$ transformation \citep{sprenger+21}, which can be used for precise arc curvature measurements \citep{baker+22, sprenger+22}. This will be left to future work.

\subsubsection{PSR J1600$-$3053}

Similar to PSR J0613$-$0200, this source shows compact features, with power extending at times to $\sim 16\,\upmu$s in the $\tau-$axis of the secondary spectrum.  The qualitative behaviour of the arcs changes over the course of our observations. In 2016, arcs appear rather broad and diffuse compared to other years, suggesting the combined contributions of multiple screens, or a larger degree of isotropy of the primary screen. In 2019$-$2020, the secondary spectra are dominated by a small number of discrete moving features, and in observations from 2020-07-25 onwards, there is only a small, featureless concentration of power at low $\tau \lesssim 1\,\upmu$s.
The total time delays were variable around a mean value of $\sim 400$\,ns, decreasing to $<200$\,ns between 2019 and 2021.  Variable scattering of this source at L-band was also found in analyses of the PTA pulsar noise contributions from timing
\citep{goncharov+21b, nanograv12_wideband, chalumeau+22}
The nature of the secondary spectra meant that precise arc curvature measurements were difficult for most observations; along with complications from potential changes in the properties of the screen, we leave modelling of the variable arc curvatures of this source to future work.

\subsubsection{PSR J1643$-$1224}

The annual and orbital variations of this pulsar were previously studied in \citet{mall+22}, placing the dominant screen distance coincident with Sh 2-27, a large diameter foreground HII region. 

PSR J1643$-$1224 was regularly observed with LEAP from 2012$-$2018, and is observed with Effelsberg as part of EPTA observations, extending our dataset from 2012-2022. 
The properties of scintillation arcs in the last year of EPTA observations (i.e. both $\eta$ and their extent in $\tau$) are still consistent with the previous trends, suggesting long-term stability of the screen(s). While the annual variations of the arcs and thus the screen geometry have been stable for $\sim 10$ years, the time delays $\langle \tau \rangle$ are seen to be variable. The scattering measured by scintillation decreases from $\sim 5 - 2.5\,\upmu$s, and roughly correlates with the decreasing DM of the source \citep{nanograv20a}.  The scintillation arcs show a persistent asymmetry which is likely related to the DM gradient in this system, which we discuss in Section \ref{sec:phasegrad}.  

We note that this source has the finest scintles of all of the LEAP sources, with $\nu_s \sim 100$\,kHz.  At the highest time delays, the scale of the scintillation pattern on Earth is $\sim 2500$\, km, nearing the length of LEAP's longest baselines. Equivalently, the angle corresponding to the largest delays is $\theta \sim 18\,$mas, while the resolution of LEAP's longest baselines is $\lambda/D \approx 21\rm{cm}/1200\,$km $\approx 35$\,mas.  For a pulsar scattered much beyond this, the angle $\theta$ of the furthest images on the sky will be outside of the LEAP beam and be resolved out during coherent addition, complicating the use of LEAP as a single effective telescope.
By the same effect, pulsars with scattering comparable to PSR J1643$-$1224 can have their screens imaged through VLBI, providing an independent way to determine scattering screen parameters \citep{brisken+10}.  Indeed, \citet{ding+23} measure PSR J1643$-$1224 to be angularly broadened to $\theta = 3.65 \pm 0.43$\,mas using the Very Large Baseline Array, and confirm the association of the dominant scattering screen with Sh 2-27.                                                                               

\subsubsection{PSR J1918$-$0642}

The scintillation timescale is $\sim$10 minutes at its shortest,
sufficiently short to reveal a faint arc in LEAP observations. At its slowest, the scintillation timescale is greater than the observation length, appearing in the secondary spectrum as power along the $f_{\rm D}=0$ axis.  Moreover, the source changed significantly around 2016, as scintles beforehand were tens of MHz, and transitioned to $\lesssim 1$ MHz afterwards, indicating a large rise in scattering time (see \ref{fig:J1918-0642secspecs}).  Coincidentally, LEAP observations taken until January 2016 were only 20 minutes long, insufficient to resolve scintillation in time.  As the arcs were difficult to resolve, we have too few curvatures measurements to fit for annual and orbital variations
but we do note that the arc curvature is clearly variable with contribution from both, as it is not always the same at a given time of year or orbital phase.

\section{Discussion}
\label{sec:discussion}

\subsection{Time Delays}
\label{sec:timedelays}
The measured values of $\langle \tau \rangle$ are shown in Figure \ref{fig:timedelays}, with a summary of results in Table \ref{tab:measurements}. The precision of pulse times-of-arrival is $\sim 1 \upmu$s in the most precisely-timed EPTA sources \citealt{chen+21}.  While the scattering timescales are less than this for most of the sources in our sample, 
uncorrected scattering variations could be a significant source of red-noise, as they are correlated in time \citep{ goncharov+21}.  Moreover, large scale variations in $\langle \tau \rangle$ are correlated on the timescale of years in several sources, and variations on similar timescales could masquerade as a common signal shared between pulsars. The subset of pulsars with resolvable scintillation with LEAP have the longest scattering times by design, 
but this subset also includes some of the pulsars with the highest timing precision in the EPTA.  In particular, pulsars J0613$-$0200, J1600$-$3053, and J1918$-$0642 are among the top 5 most significant contributors to the common red-noise process detected by the IPTA \citep{antoniadis+22}, and all show variable scattering on $\gtrsim 100\,$ns level in this work.  As we approach a potential GW detection, and as PTA sensitivity increases, the variable time delays from scattering will be important to consider in GW searches.

\subsection{Annual and Orbital Fitting of PSR J0613$-$0200}
\label{sec:annualorbital}

In \citet{main+20}, a strong annual trend in the arc curvatures of PSR J0613$-$0200 was seen. Fitting the annual variations resulted in a fractional screen distance of $s=0.58\pm0.10$ during the period of increased scattering in 2013, and $s=0.62 \pm 0.06$ afterwards, with screen axis $\psi$ changing by $\sim 50^{\circ}$. It was argued that the scattering could originate in the same screen, with orientation changing over time.
The orbital variations were ignored, and were an additional source of scatter in individual measurements.

Here, we revisit the curvature variations of PSR J0613$-$0200, including the most recent data, improved measurement of arc curvatures, and including fitting orbital variations using the same framework as in \citet{mall+22}. We use Gaussian priors on the proper motion and distance from the most recent IPTA values, $\mu_{\alpha} = 1.828(5)$mas/yr,  $\mu_{\delta} = -10.35(1)$\,mas/yr,  $d_{\rm psr} = 1.11\pm0.05$\,kpc \citep{perera+19}.  
We fit the distance weighted effective velocities $W$ with a model of an anisotropic scattering screen, 
\begin{align}
    W = \frac{1}{\sqrt{d_{\rm eff}}} \bigl|&\left( \frac{1-s}{s} v_{\rm psr, \alpha} + v_{\earth, \rm \alpha} - \frac{1}{s}v_{\rm scr, \alpha} \right)\sin(\psi) + \\
    &\left( \frac{1-s}{s} v_{\rm psr, \delta} + v_{\earth, \rm \delta} - \frac{1}{s}v_{\rm scr, \delta} \right)\cos(\psi)\ \bigr| \nonumber.
\end{align}
We compare several different models, detailed in the following sections.

\subsubsection{Variations in Screen Properties}
From \citet{main+20}, we know a single screen is a poor fit to the full 2013-2019 dataspan. Here, we try two models to account for the variable screen.  For each model, we allow for $N$ jumps of the screen parameters, where the times of the jumps are free parameters, bounded between the time of the first and last observation. In the first model, we assume that the scattering originates from a single screen, which only changes in orientation over time (similar to the 1D scattering screens of PSR B0834+06, and B1508+55 \citep{simard+19, sprenger+22}.  In this case, the fractional screen distance $s$, and 2D velocity $v_{\rm scr, \alpha}$, $v_{\rm scr, \delta}$ are free parameters and constant over the full duration. In the second model, we allow $s$, $\psi$, and $v_{\psi}$ to vary in each jump.  This is highly similar to the approach in \citet{walker+22}, who model the arc curvatures of PSR J1603$-$7202. They allow all screen properties to change between jumps, and find moderately strong support for 2 jumps, one corresponding to a region of enhanced DM and scattering.

In addition, we fit for the orbital inclination $i$ and angle of nodes $\Omega$, and we include white noise parameters F and Q as free parameters of the fit, such that the scaled errors are $\delta W_{\rm corr} = \sqrt{ (\rm{F\, \times\,\delta W})^{2} + \rm{Q}^{2} }$.  These parameters can account for biases and underestimated errors, but could also arise physically from variations in screen axis and velocity, which could vary on the refractive timescale $t_{r}$ (e.g. \citealt{askew+23}). 


The first model, allowing for all screen parameters to vary in each jump, results in $BIC=483.4$, while the second model with a screen at a fixed distance, changing only the orientation has $BIC=442.5$.  Both models have almost identical white noise parameters $F$ and $Q$, suggesting that they fit the data comparably well, but the first model is penalized for having more free parameters.  We suggest that a single screen can reproduce our observations, but we cannot rule out the possibility that the variable screen properties correspond to different screens dominating at different times.  The best fit values of these models are tabulated in Table \ref{tab:J0613fittable}.  

\subsubsection{Orbital Constraints}

Scintillation arcs provide a way to measure resolve the ambiguity in the sense of the inclination, i.e. $i < 90^{\circ}$, or $i > 90^{\circ}$.  We obtain a fit orbital inclination of $i = 58\pm4^{\circ}$ is consistent with one value from timing of $i_{\rm timing} = 68^{\circ^{+7}}_{-10}$ \citep{fonseca+16}, but mildly inconsistent with the IPTA value of $i_{\rm timing} = 70\pm3^{\circ}$ by $< 2\sigma$.  We obtain the first measurement of $\Omega =124\pm4^{\circ}$.   Additionally, we compare to the fit restricting $i>90$.  This results in values of $i=116.4 \pm 8.4$, and $\Omega = 274.0 \pm 12.8$, but is disfavoured, with $BIC=523$

\begin{figure}
\centering
\includegraphics[width=1.0\columnwidth, clip=true]{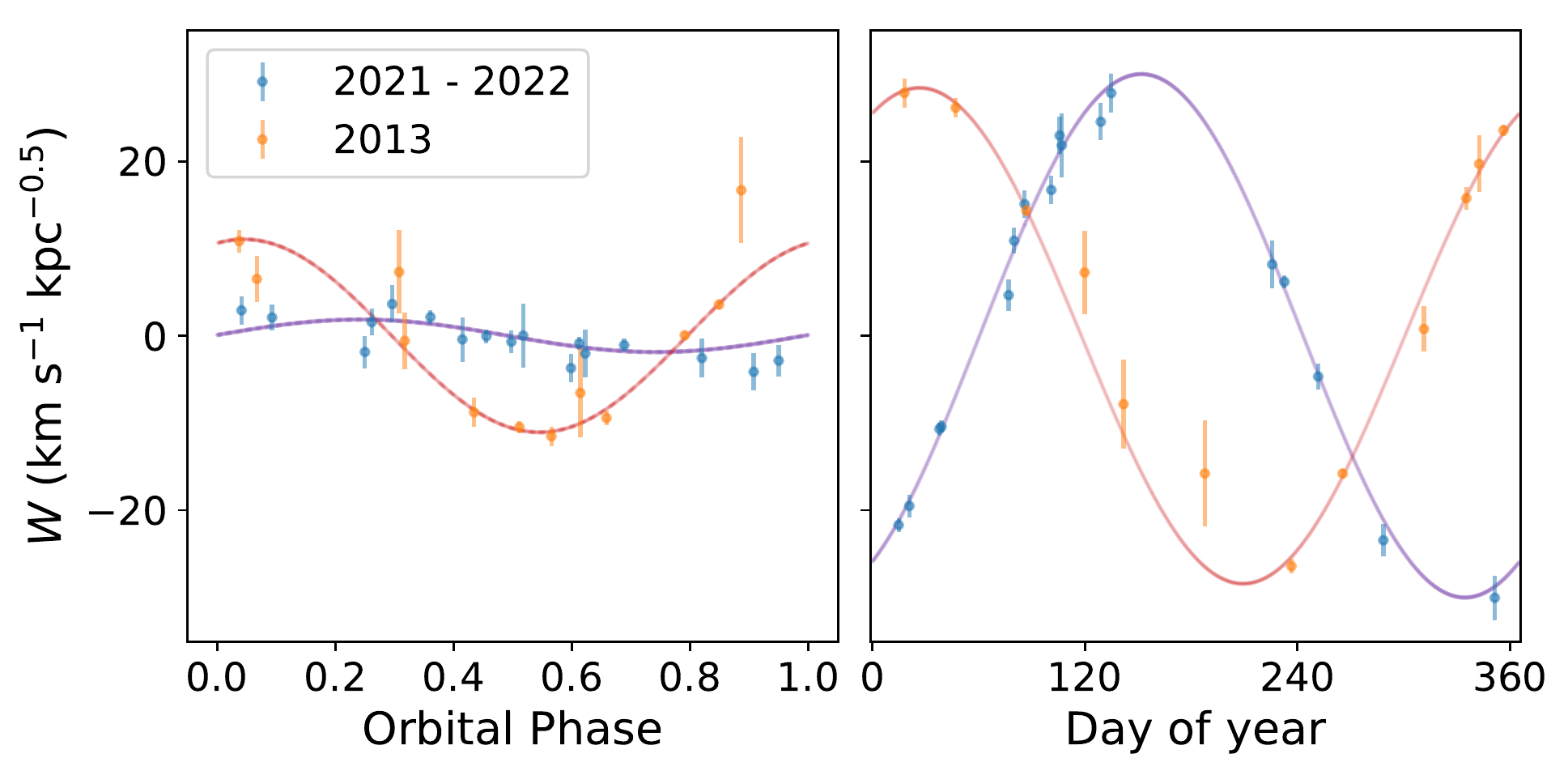}\\
\vspace{-5pt}
\caption{Results of modelling annual and orbital scintillation arc variations of PSR J0613$-$0200, shown during two periods of heightened scattering where arcs could be measured precisely.  The orbital amplitude, and the phase of the annual curves are clearly different between the two, indicating a change of screen geometry; while $\Omega$ and $\psi$ are almost perpendicular in 2021$-$2022, the changing screen orientation ensures that the orbital modulation is seen.}
\label{fig:J0613-0200annorb}
\end{figure}

The best fit model is overlaid on the timeseries of $W$ measurements in Figure \ref{fig:timedelays}, and the decomposition to annual and orbital velocity is shown in Figure \ref{fig:J0613-0200annorb}. The variation of the properties of the observed screen
can be clearly seen as the phase of the annual maxima and minima change over time, and the amplitude of the orbital curve changes due to the changing alignment of $\psi$ and $\Omega$.  The times of the jumps both correspond to regions of changing $\langle \tau \rangle$, suggesting that both of these effects trace physical changes of the screen.

\subsubsection{Comparison to previous work}

Our results are largely consistent with, yet more precise than \citet{main+20}.  This is unsurprising, as both analysis contain much of the same data, but the measurements of $W$ are made differently, and our present analysis includes data beyond 2020, orbital variations, and times of screen jumps as free parameters rather than fixed.
However, our value of $s=0.71\pm0.02$ is smaller than the value of $s=0.62\pm0.06$.  This is a result of the different pulsar distance used, $d_{\rm psr}=780\pm80\,$pc from \citet{desvignes+16}, and both measurements result in a consistent screen distance. 

\begin{figure}
\centering
\includegraphics[width=0.8\columnwidth, clip=true]{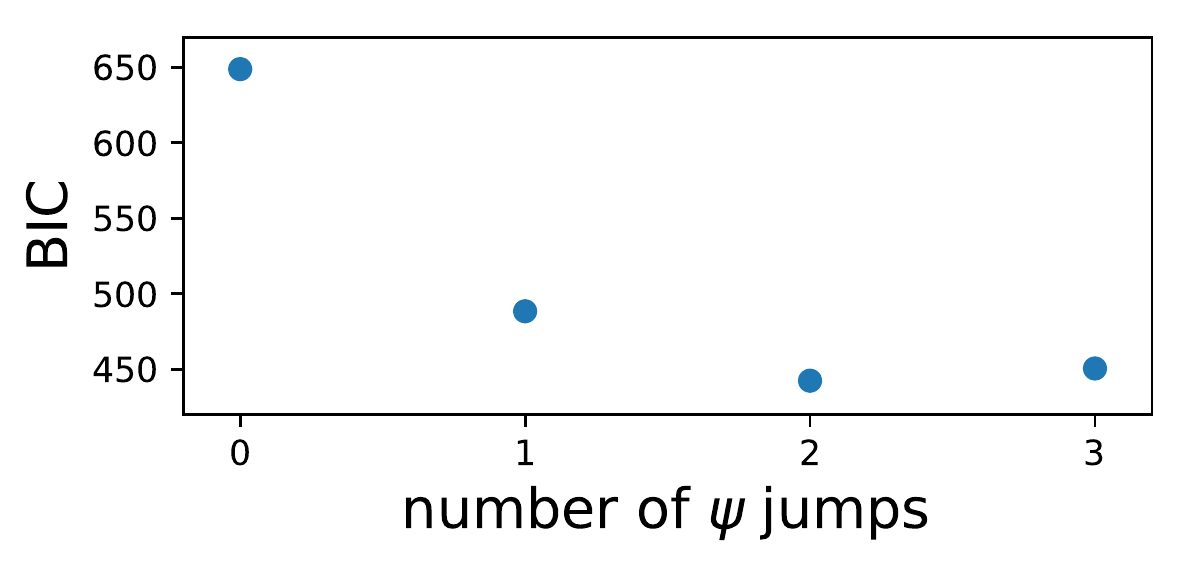}\\
\vspace{-5pt}
\caption{Bayesian Information Criteria 
in fitting $W$ variations of PSR J0613$-$0200, for models with $N$ jumps in the screen orientation $\psi$. Models with fewer than 2 jumps result in a poor fit, while models with more than than 2 jumps result in no additional improvement and result in increased BIC due to having more parameters.}
\label{fig:J0613redchi}
\end{figure}

\begin{table}
    \centering
    \caption{Results of modelling arc curvature variations of PSR J0613$-$0200, and comparison to \protect{\citet{main+20}}.  The model assumes an anisotropic scattering screen at a fixed distance, but allowing for 2 jumps in the screen orientation $\psi$, as described in Section \protect{\ref{sec:annualorbital}}. The top set of parameters are the free parameters of the model, while the bottom set includes the pulsar distance prior (distances from $*:$ \protect{\citet{perera+19}}, $\dagger:$ \citet{desvignes+16} ), and derived quantities.}
    \begin{tabular}{cccc}
    \hline
    Parameters & Model 1 & Model 2 &  Main et al. 2020 \\
    \hline
    $s$ & $0.71^{*}\pm0.02$ & -  & - \\
    $s0$ & - & $0.40^{*}\pm0.19$  & $0.58^{\dagger}\pm0.10$ \\
    $s1$ & - & $0.73^{*}\pm0.03$ &     $0.62^{\dagger}\pm0.06$ \\
    $s2$ & - & $0.50^{*}\pm0.12$ & - \\
    $v_{\rm{scr},\alpha}$ (km s$^{-1}$) &  $17.2 \pm 0.9$ & -  &  - \\
    $v_{\rm{scr},\delta}$ (km s$^{-1}$) & $-4.4 \pm 0.8$ & - & - \\
    $\psi_{0}(^{\circ}$) &  $-30 \pm 3$  & $-31\pm3$  & $-36 \pm 9$  \\
    $\psi_{1}(^{\circ}$) &  $16 \pm 2$  & $16\pm2$ & $16 \pm 2$ \\
    $\psi_{2}(^{\circ}$) &  $40 \pm 3$  & $42\pm3$  & - \\
    $i (^{\circ}$) & $58 \pm 4$  & $55 \pm 5$ & -\\
    $\Omega (^{\circ}$) & $124 \pm 4$  & $126 \pm 5$ & -\\ 
    $T_{1, \rm jump}$ (mjd) & $56670\pm30$ & $56667 \pm 30$ &  $56658$ (fixed) \\
    $T_{2, \rm jump}$ (mjd) & $58820\pm30$ & $58820\pm25$ &  - \\
    $Q$ & $4.4 \pm 0.9$ & $4.4 \pm 1.0$ &  - \\
    $F$ & $3.1 \pm 0.9$ & $3.3 \pm 0.9$ & - \\
    \hline
    $d_{\rm psr}$ (kpc) & $1.11^{*}\pm0.05$ & $1.12^{*}\pm0.05$ & $0.78^{\dagger}\pm0.08$ \\
    $d_{\rm scr}$ (kpc) & $0.32\pm0.04$ & $0.30\pm0.03$ &  $0.30\pm0.07$ \\
    $v_{\psi_0}$ (km s$^{-1}$) & $0.5 \pm 0.8$ & $2.1\pm1.9$ &  $-1.2\pm2.5$  \\
    $v_{\psi_1}$ (km s$^{-1}$) & $12.4 \pm 0.8$ & $11.3 \pm 1.2$  &  $12.8\pm2.8$  \\
    $v_{\psi_2}$ (km s$^{-1}$) & $7.7 \pm 0.9$ & $10.7 \pm 1.3$ &  - \\
    BIC  & 455.8  & 470.5  & - \\
    \hline
     \end{tabular}
     \label{tab:J0613fittable}
     \newline
\end{table}

\subsection{Movement of Features}
\label{sec:featuremovement}

Several of the sources in our sample show discrete, compact regions of power in their secondary spectra. as described in Section \ref{sec:timescale}, the movement of compact features through the secondary spectrum can be predicted through the arc curvature.  We investigate the feature movement in PSR J0613$-$0200 and PSR J1600$-$3053.  

\begin{figure}
\centering
\includegraphics[width=1.0\columnwidth, trim=0.32cm 0cm 0.35cm 0.0cm, clip=true]{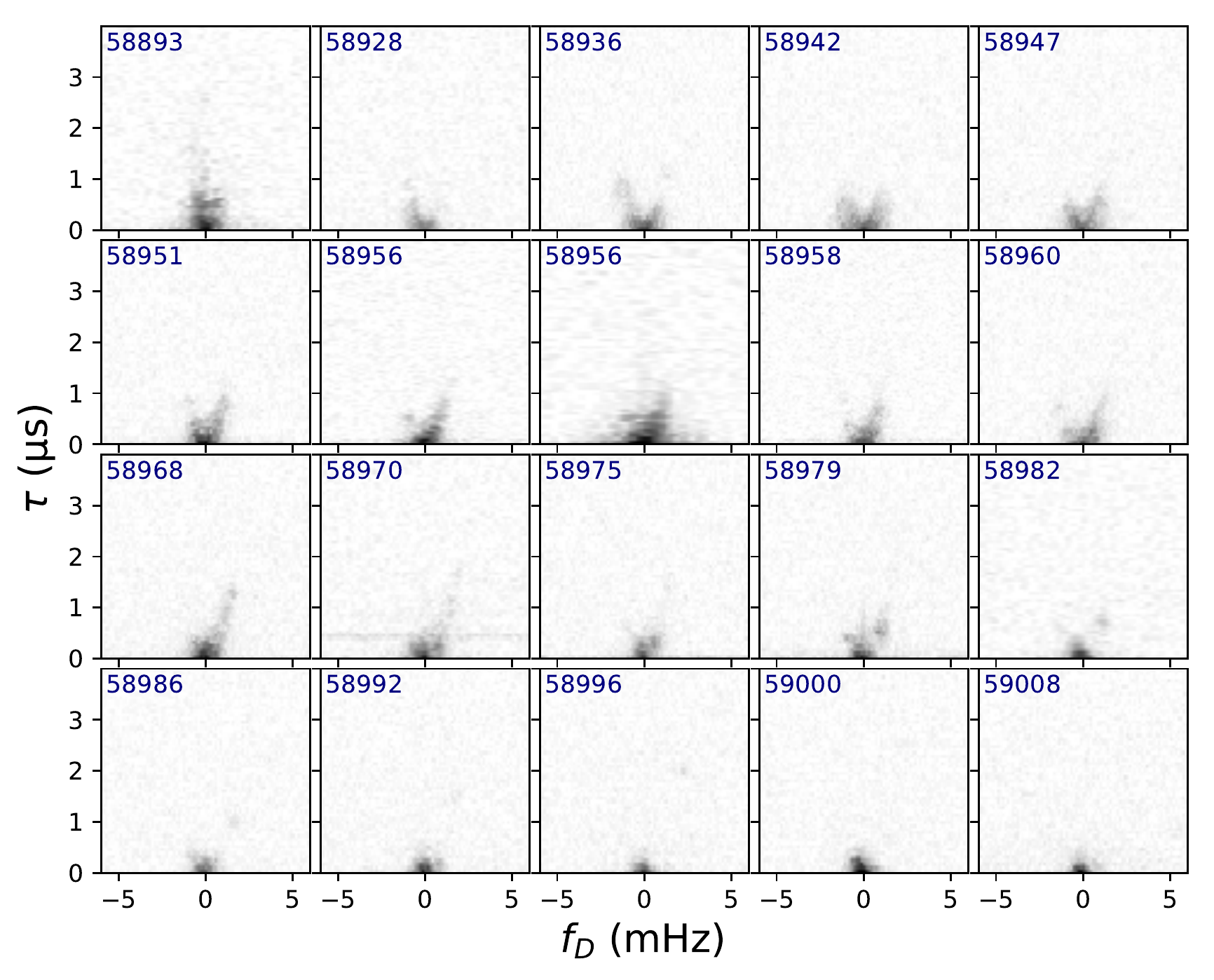} \\ 
\hspace{7pt}
\includegraphics[width=0.96\columnwidth, trim=-0.2cm 0cm 0.35cm 0.8cm, clip=true]{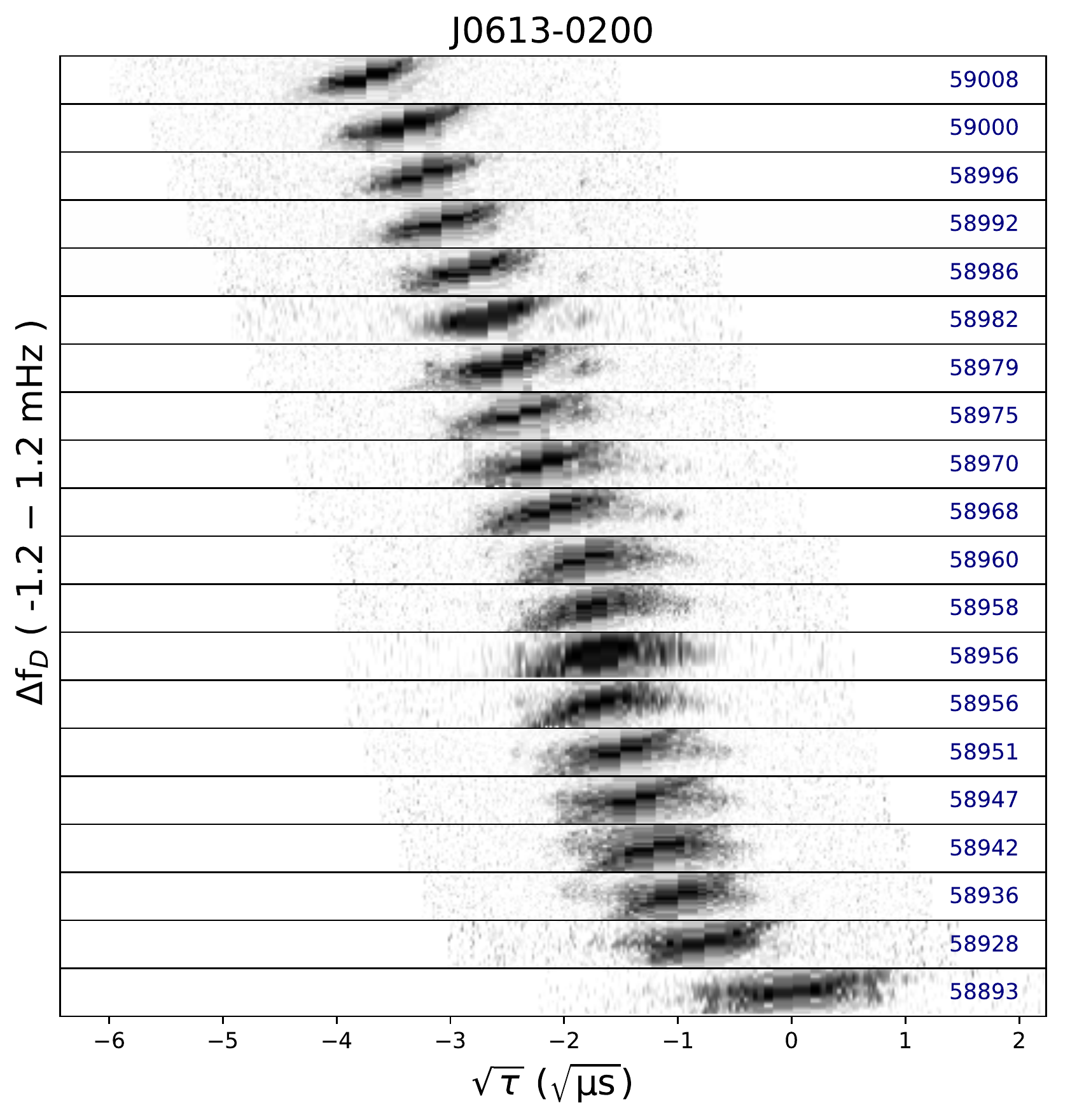} \\
\caption{ Feature alignment in PSR J0613$-$0200, during the dense Effelsberg observing campaign. \textit{Top:} Secondary spectra, where several discrete features can be seen to move throughout. \textit{Bottom:} Profiles of $\sqrt{\tau}$ vs.~$f_{\rm D}$ from the corresponding secondary spectra, made as described in Section \protect{\ref{sec:featuremovement}}, and shifted by the predicted movement between observations. The value $\sqrt{\tau} \propto \theta$ is a proxy for the image positions; features connected vertically between observations suggest persistent scattering at regions of fixed $\theta$.}
\label{fig:J0613features}
\end{figure}

\begin{figure}
\centering
\includegraphics[width=1.0\columnwidth, trim=0.32cm 0cm 0.35cm 0.0cm, clip=true]{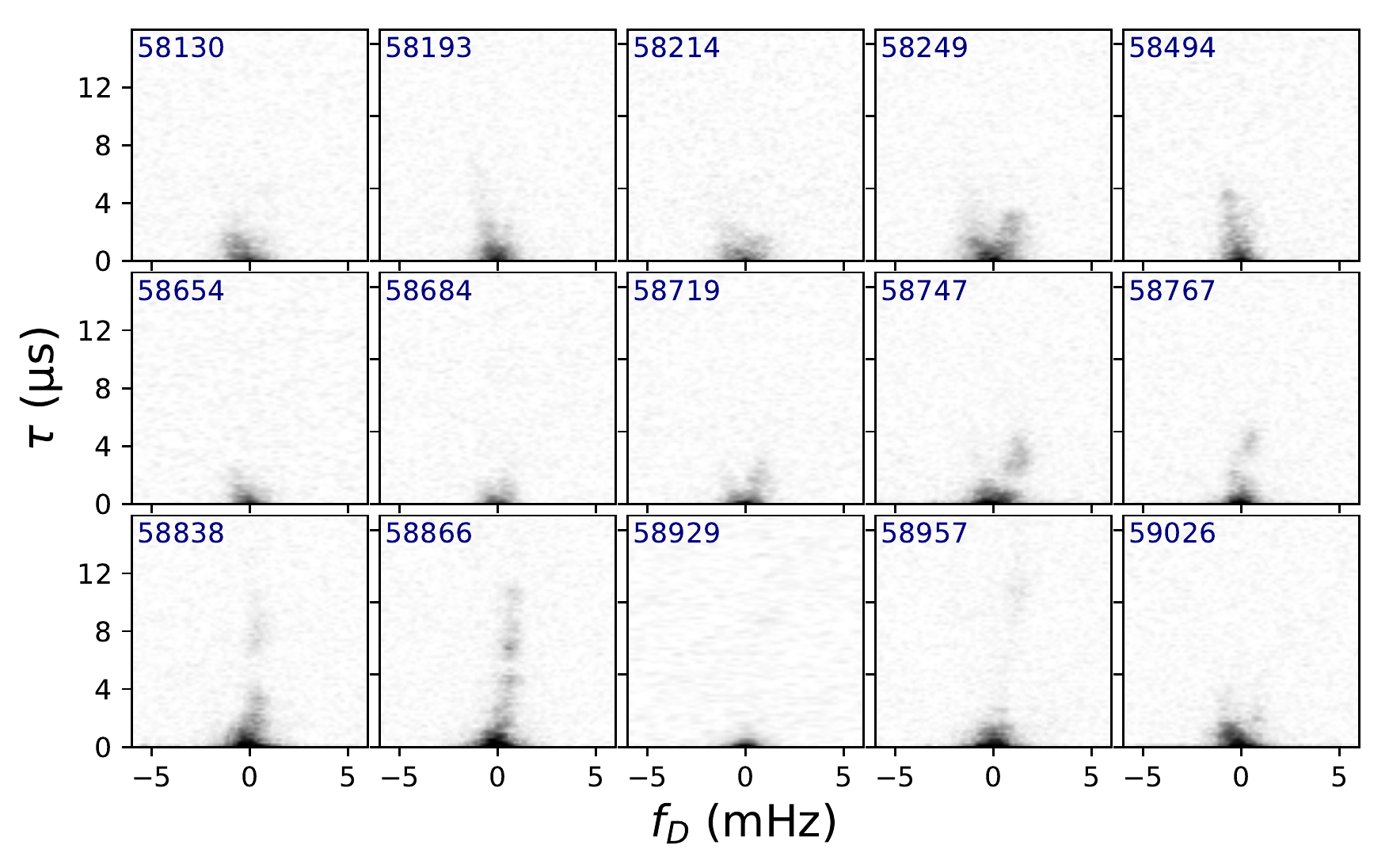} \\ 
\hspace{7pt}
\includegraphics[width=0.95\columnwidth,  trim=-0.2cm 0cm 0.35cm 0.8cm, clip=true]{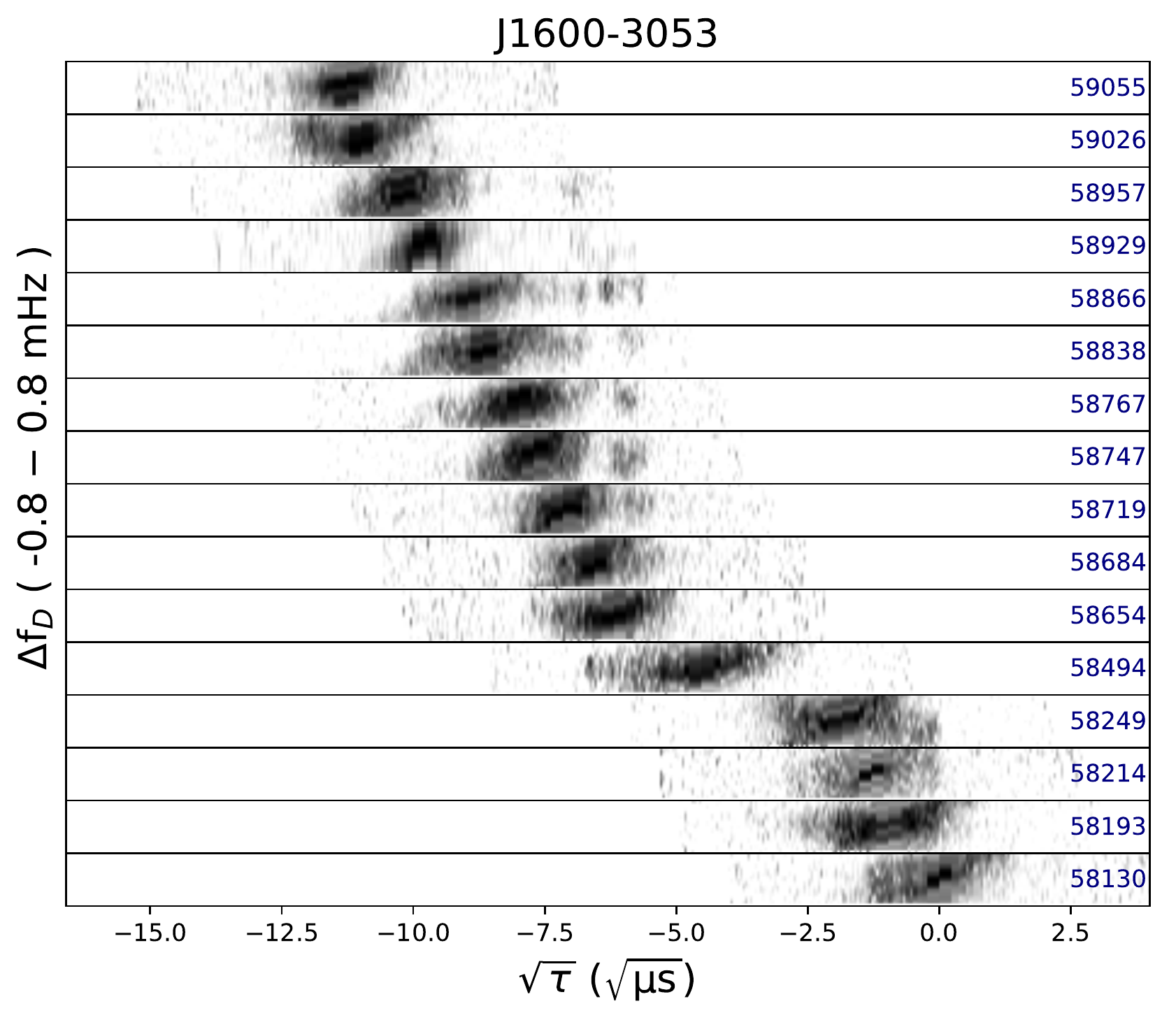} \\
\caption{ Same as \ref{fig:J0613features}, but for a series of LEAP PSR J1600$-$3053 observations showing moving features.}
\label{fig:J1600features}
\end{figure}

Following the techniques of \citet{sprenger+22}, we remap the secondary spectra in terms of $\sqrt{\tau}\propto \theta$.  Surrounding the best fit parabola $\eta$ of each secondary spectrum, we take a slice of $I(f_{\rm D}, \tau_{i})$ for each bin in $\tau$, remapping every $\tau$ bin to the closest value of $\sqrt{\tau}$ to form a profile of $I(f_{\rm D}, \sqrt{\tau})$.  The shift of features between observations is predicted by the annual model of $\eta$ (assuming the orbital variations average out over time).  The results of a subset of the data in PSR J0613$-$0200 and PSR J1600$-$3053 is shown in Figure \ref{fig:J0613features} and \ref{fig:J1600features} respectively.  In both cases, features can be seen to persist at a fixed location over the time they traverse the secondary spectra, indicating scattering from compact regions of fixed $\theta$.  This can only be seen if there is significant power near the undeflected image of the pulsar; we see that there is persistently significant power surrounding $\theta = 0$ which is difficult to track between observations. This contains the bulk of the power, dominating the changing values of $\langle \tau \rangle$.

\subsection{DM Gradients and Asymmetric Arcs}
\label{sec:phasegrad}

Scintillation arcs often show a clear asymmetry in power, related to the phase structure across the scattering screen.
A local linear DM slope along the direction of $\boldsymbol{v}_{\rm eff}$ creates a refractive shift \citep{cordes+06, rickett+14}, 
\begin{equation}
    \theta_{r} = \frac{\lambda^{2} r_{\rm e}} {2\pi v_{\rm eff, \shortparallel}} \partial_{t} \textrm{DM}
\end{equation}
leading to a new zero-point of a secondary spectrum offset by $\tau_{r}$ as:
\begin{equation}
    \partial_{t} \textrm{DM} = \frac{2\pi v_{\rm eff, \shortparallel}}{\lambda^{2} r_{\rm e}} \sqrt{ \frac{2c\tau_r}{d_{\rm eff}} }.
\end{equation}
Under these assumptions, the gradient in DM within the screen can be estimated from the asymmetry in scintillation arcs, and vice versa (shown in practice in 
\citet{reardon+23} using scintillation ACFs). Additionally, the relation between the two depends on the distance weighted effective velocity; connecting all related observables will allow for the maximum amount of information to be extracted about intervening scattering screens.

\begin{figure}
\centering
\includegraphics[width=1.0\columnwidth, clip=true]{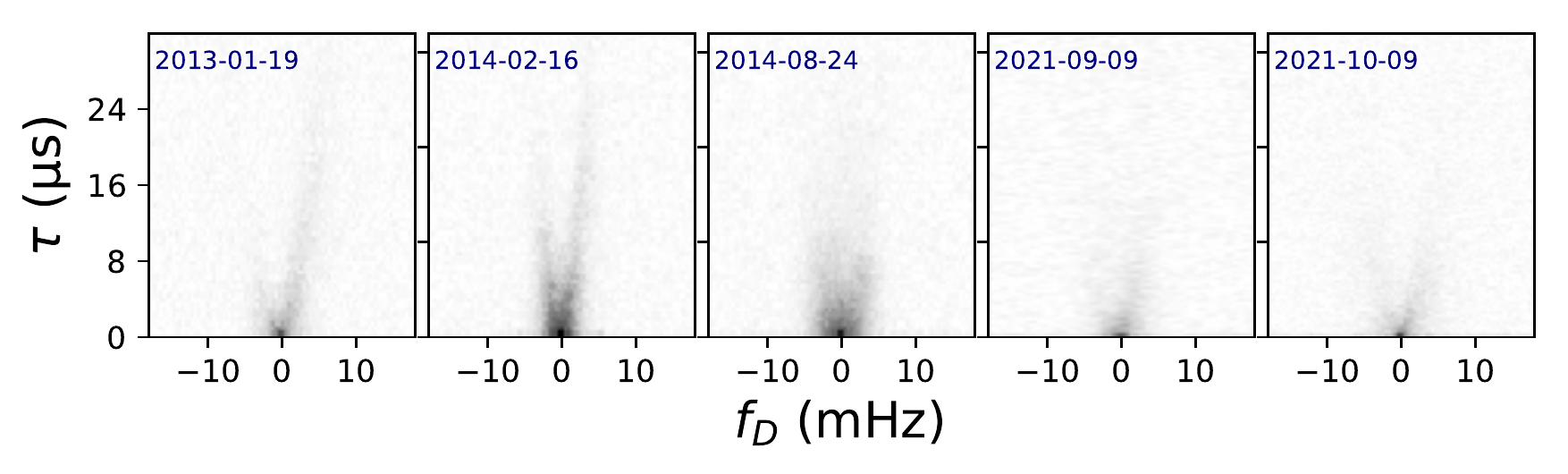}\\
\includegraphics[width=1.0\columnwidth, clip=true]{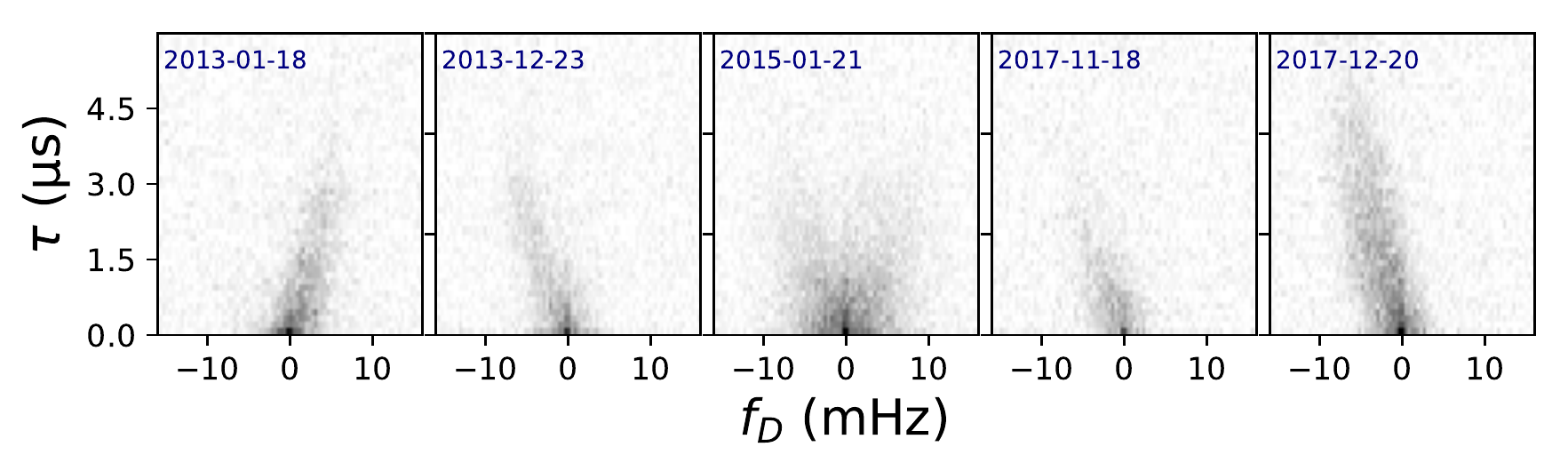}\\
\vspace{-5pt}
\caption{Top: PSR J1643$-$1224 secondary spectra at a similar time of year, showing a very similar distribution of power, with an asymmetric arc with power at positive $f_{\rm D}$.  The persistent asymmetric power distribution suggests decreasing DM along $\boldsymbol{v}_{\rm eff}$, and thus decreasing DM with time. Bottom: PSR J0621+1002 secondary spectra, showing diffuse, highly asymmetric power of changing signs.  This likely suggests variable DM in PSR J0621+1002, but could also arise from the sign of $\boldsymbol{v}_{\rm eff}$ changing from orbital motion; the orbit of PSR J0621+1002 will be investigated in future work. }
\label{fig:AsymmetricArcs}
\end{figure}

In our sample, PSRs J0621+1002 and J1643$-$1224 are the clearest examples showing diffuse, highly asymmetric scintillation arcs, likely reflecting significant variations in DM (example secondary spectra shown in Figure \ref{fig:AsymmetricArcs}). 
The DM curve of PSR J1643$-$1224 from the NANOGrav 12.5 year data release shows a persistent downwards trend in DM of $\Delta \rm{DM} \sim 10^{-3}$\dmu $\rm{year}^{-1}$ from $\sim 2013-2016$ \citep{nanograv20a}, during which time the scintillation arcs showed persistent asymmetric power to the right quadrant of the secondary spectrum. The sign of asymmetry in PSR J0621+1002 changes on the timescale of months, which may suggest rapidly varying DM.  This explanation is plausible as observations with LOFAR at frequencies of about 140\,MHz the DM of J0621+1002 has been seen to vary by $\sim 10^{-2}$\dmu on several month timescales \citep{donner+20}.  However, we note that the scintillation arcs are sensitive to DM gradients within the scattering screen, not necessarily the total changing electron column which is measured by timing.
A detailed analysis comparing high-cadence DM and scintillation arc asymmetries will be valuable, but is beyond the scope of this paper.

\section{Conclusions and Future Prospects}
\label{sec:conclusions}

In this paper, we performed the first large-sample study of scintillation arcs in MSPs, where of 22 sources regularly observed at LEAP, we observed scintillation arcs in 12.  We are able to measure the time-variable arc curvature and scattering in 6 of these sources, with $\sim$monthly cadence over $5-10$\,years.

The scintillation arcs reveal the structure along the dominant scattering screens in these sources, revealing varying phenomena, including compact sources of scattering in PSRs J0613$-$0200 and J1600$-$3053, asymmetric distributions of power likely reflecting DM gradients in PSRs J0621+1002 and PSR J1643$-$1224, and multiple arcs indicating scattering by multiple thin screens along the line of sight in PSR B1937+21.  In fitting of the variable scintillation arc curvatures of PSR J0613$-$0200, we were able to measure $\Omega$, and resolve the sense of $i$, finding a value of $i$ consistent with pulsar timing.  The screen axis of PSR J0613$-$0200 changes by tens of degrees over 10 years ($\sim 100\,$AU), corresponding to visible changes in the extent of scattering.

The time delays measured through scintillation can be compared and combined with other methods, including scattering measured at lower frequencies, measured through sharp features such as giant pulses \citep{bilous+15, main+17, mckee+19} or microstructure \citep{liu+22}.  The effects of correlated, variable scattering, as well as correction methods will be assessed using simulations and applied to PTA data in future work. 

Orbital studies using scintillation can be improved with better understanding of scattering screens, and with more precise measurements of the arc curvature. Studies to date have all been incoherent, attempting to measure the primary scintillation arc without full information of inverted arclets results from interfering pairs of images which arise in highly anisotropic screens.  Phase retrieval techniques such as holography \citep{walker+08, oslowski+23}, cyclic spectroscopy \citep{demorest11, walker+13}, and the $\theta-\theta$ transformation \citep{sprenger+21, baker+22} can greatly increase the precision.  In sources with discrete features, the movement of features between observations gives another constraint on the average arc curvature in the time between observations, and can be used as an additional precise constraint \citep{sprenger+22}.  Even without these advanced techniques, improved cadence of observations offer a great improvement, to better fill the annual and orbital planes, to track features between observations, and to track screen changes.  Using all available data, including measurements of scintillation velocities in conjunction with arcs, will result in better constraints on pulsar orbits and screens.

Much larger than the effect of scattering variations are the changes in DM, for which there is significant effort to measure (e.g. \citealt{jones+17, donner+20, chimepulsar21, inpta22}).  Scintillation, scattering, and refractive flux variations are all physically linked, and related to the changing column density of electrons.  Detailed mappings of these quantities, as is now being attempted in eclipsing binaries \citep{lin+21, lin+22}, will be valuable, and lead to a more complete physical understanding of the effects of the IISM on pulsar signals.

\section*{Acknowledgements}

We thanks Daniel Reardon for useful comments, and for advice on modelling arc curvature variations.
This work is supported by the ERC Advanced Grant ``LEAP", Grant Agreement Number 227947 (PI M.\,Kramer). 
The European Pulsar Timing Array (EPTA) is a collaboration between European Institutes, namely ASTRON (NL), INAF/Osservatorio Astronomico di Cagliari (IT), the Max-Planck-Institut f{\"u}r Radioastronomie (GER), Nan{\c c}ay/Paris Observatory (FRA), The University of Manchester (UK), The University of Birmingham (UK), The University of Cambridge (UK), and The University of Bielefeld (GER), with an aim to provide high-precision pulsar timing to work towards the direct detection of low-frequency gravitational waves.  
 
 The Effelsberg 100-m telescope is operated by the Max-Planck-Institut f{\"u}r Radioastronomie. 
 Pulsar research at the Jodrell Bank Centre for Astrophysics and the observations using the Lovell Telescope are supported by a consolidated grant from the STFC in the UK. 
The Westerbork Synthesis Radio Telescope is operated by the Netherlands Foundation for Radio Astronomy, ASTRON, with support from NWO. 
The Nan{\c c}ay Radio Observatory is operated by the Paris Observatory, associated with the French Centre National de la Recherche Scientifique. 
The Sardinia Radio Telescope (SRT) is funded by the Department of Universities and Research (MIUR), the Italian Space Agency (ASI), and the Autonomous Region of Sardinia (RAS), and is operated as a National Facility by the National Institute for Astrophysics (INAF). From Mar 2014 - Jan 2016, the SRT data were acquired as part of the Astronomical Validation of the SRT. We thus thank the SRT Astronomical Validation Team, and in particular: S. Casu, E. Egron, N. Iacolina,  A. Pellizzoni, and A. Trois.  

SC acknowledges the support by the ANR Programme d'Investissement d'Avenir (PIA) under the FIRST-TF network (ANR-10-LABX-48-01) project and the Oscillator IMP project (ANR-11-EQPX-0033-OSC-IMP), and by grants from the R\'{e}gion Bourgogne Franche Comt\'{e} intended to support the PIA. 
SC and IC acknowledge financial support from Programme National de Cosmologie and Galaxies (PNCG) and Programme National Hautes Energies (PNHE) funded by CNRS, CEA and CNES, France.
HH acknowledges the support by the Max-Planck Society as part of the “LEGACY” collaboration with the Chinese Academy of Sciences on low-frequency gravitational wave astronomy.
 JWM gratefully acknowledges support by the Natural Sciences and Engineering Research Council of Canada (NSERC), [funding reference \#CITA 490888-16].
 KL is supported by the European Research Council for the ERC Synergy Grant BlackHoleCam under contract no.\ 610058.  K.J.Lee gratefully acknowledges support from National Basic Research Program of China, 973 Program, 2015CB857101 and NSFC 11373011. DP gratefully acknowledges financial support from the research grant “iPeska” (P.I. Andrea Possenti) funded under the INAF national call PRIN-SKA/CTA with Presidential Decree 70/2016. WWZ was supported by the National Natural Science Foundation of China Grant No. 11873067 the CAS-MPG LEGACY project and the Strategic Priority Research Program of the Chinese Academy of Sciences Grant No. XDB23000000.

\section*{Data Availability}

Upon publication, the dynamic spectra will be published on zenodo at \url{10.5281/zenodo.7415215}.

\bibliographystyle{mnras}
\bibliography{leap}

\appendix
\section{Secondary Spectra}
\renewcommand{\thefigure}{A\arabic{figure}}
\setcounter{figure}{0}

Here we show the secondary spectra of all of the sources with resolvable scintillation with LEAP, as well as the Effelsberg observations of PSRs J0613$-$0200 and J1643$-$1224.

\begin{figure*}
\centering
\includegraphics[width=1.0\textwidth]{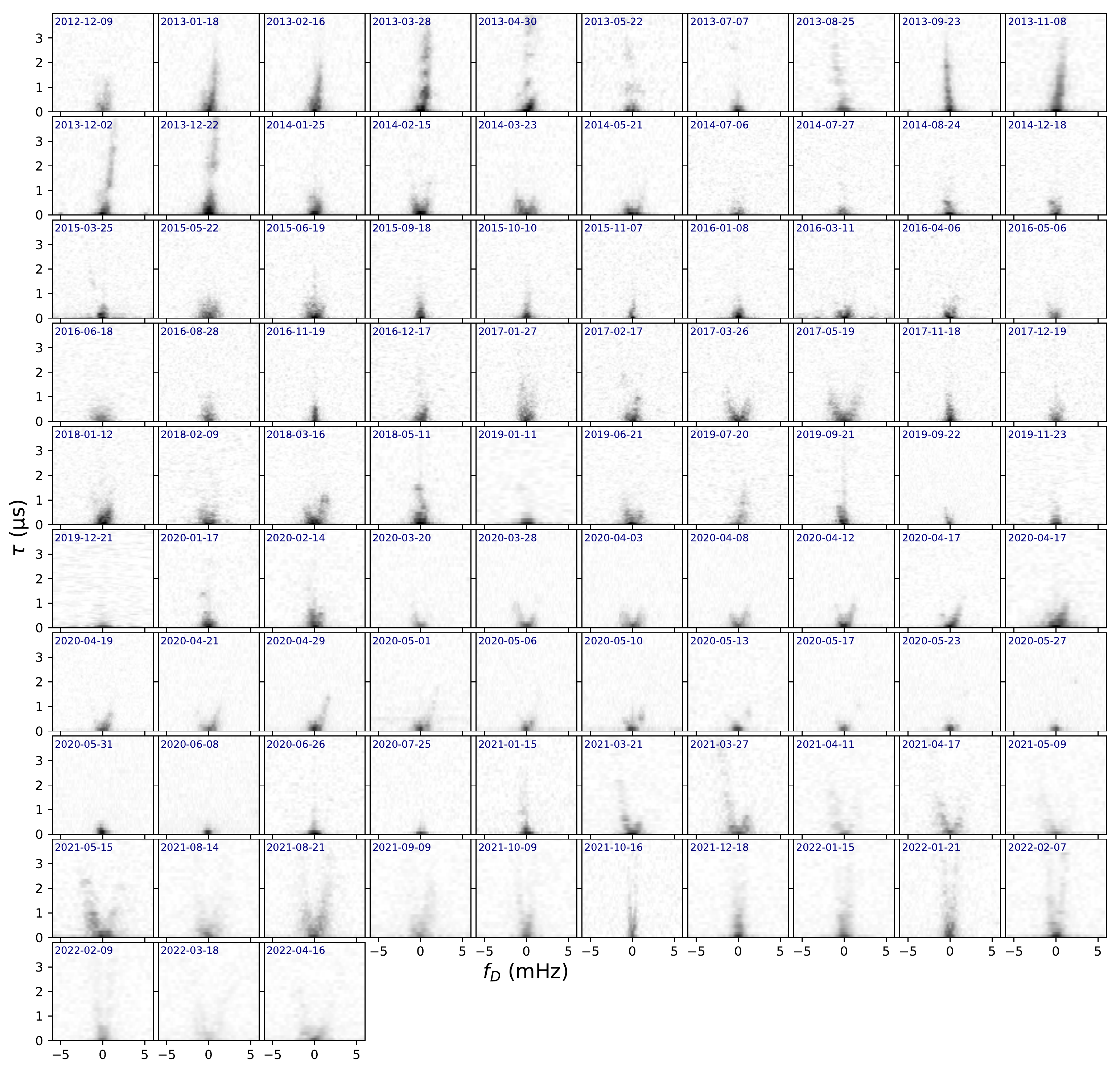}
\vspace{-0.5cm}
\caption{Secondary spectra of all observations of PSR J0613$-$0200, including targeted Effelsberg observations and recent scintillation-stream EPTA data.}
\label{fig:J0613-0200secspecs}
\end{figure*}

\begin{figure*}
\centering
\includegraphics[width=1.0\textwidth]{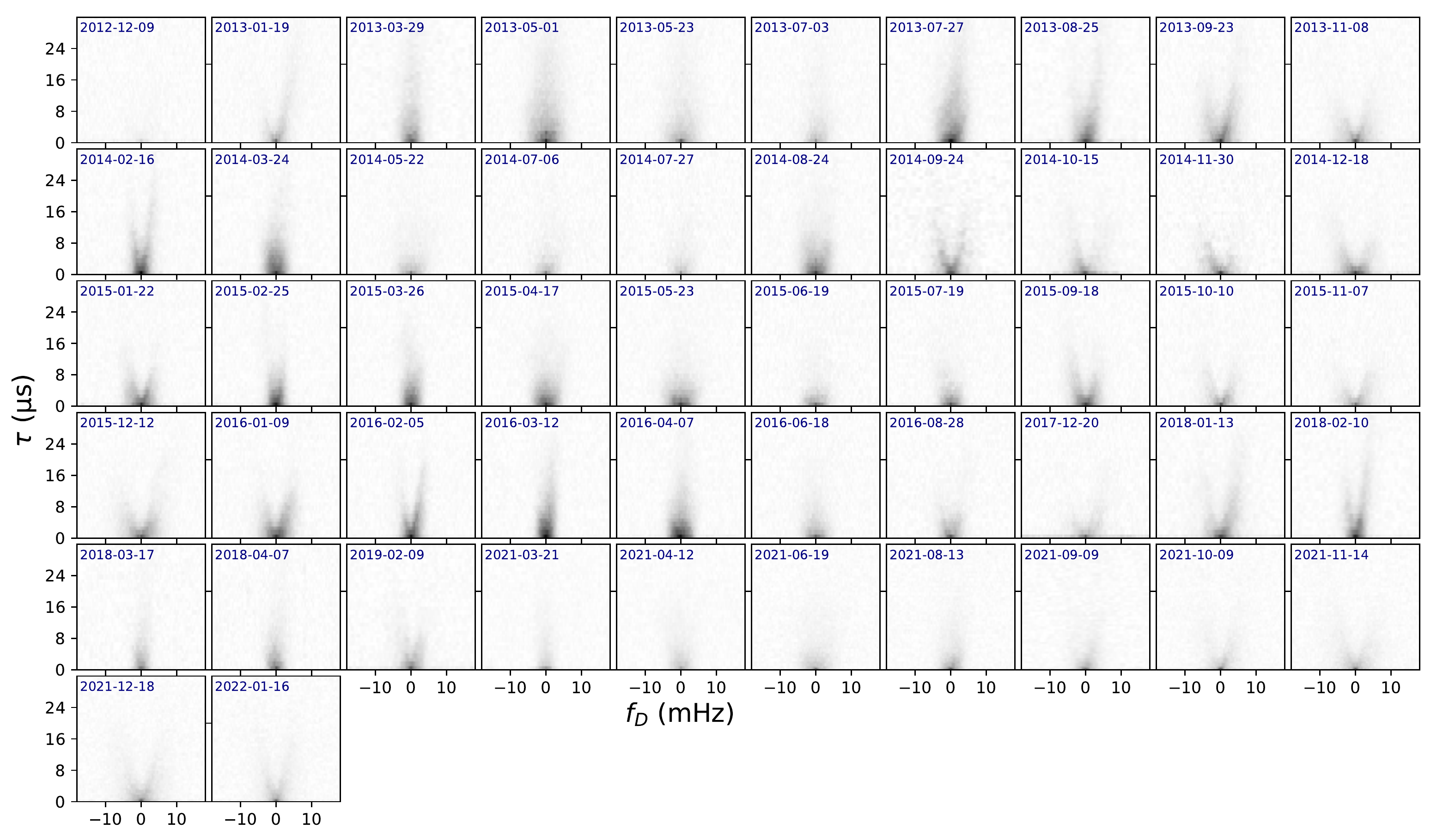}
\vspace{-0.5cm}
\caption{Secondary spectra of all observations of PSR J1643$-$1224, including recent scintillation-stream EPTA data.}
\label{fig:J1643-1224secspecs}
\end{figure*}

\begin{figure*}
\centering
\includegraphics[,width=1.0\textwidth]{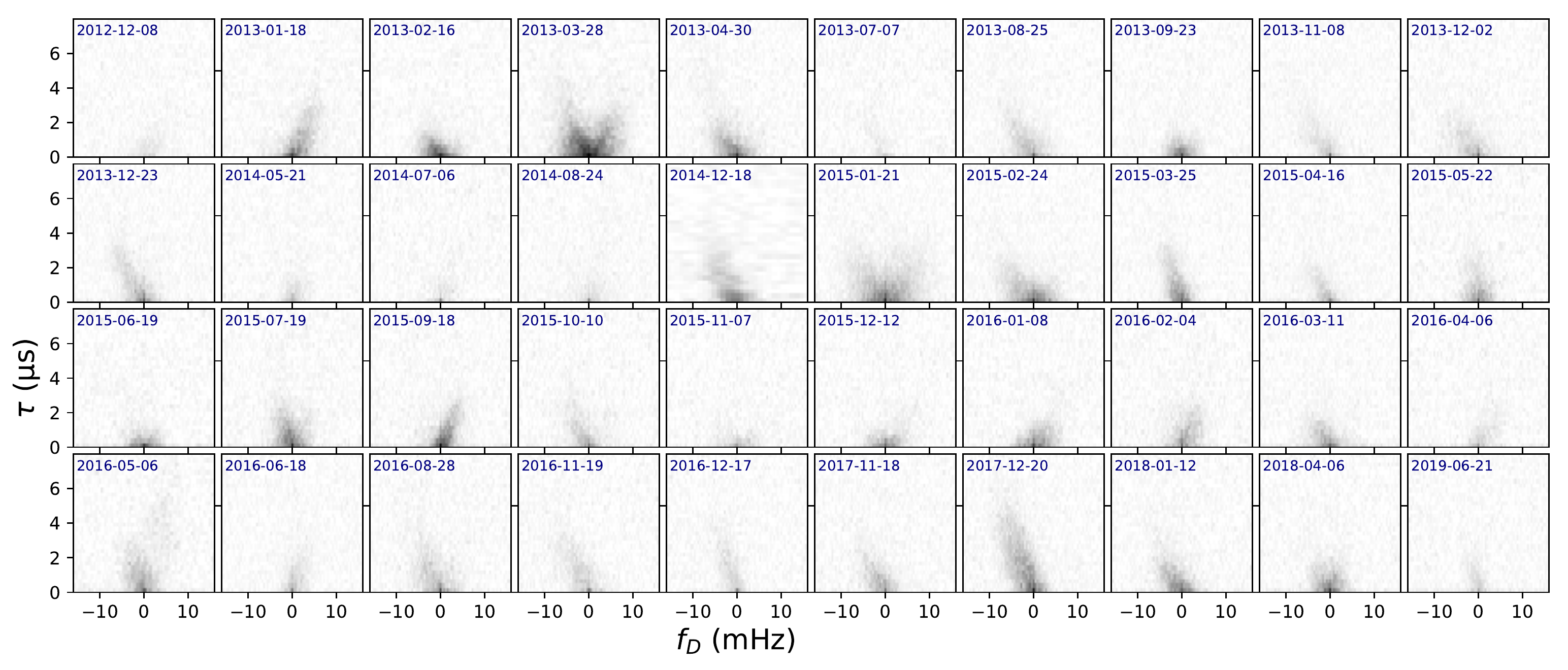}
\vspace{-0.5cm}
\caption{Secondary spectra of all observations of PSR J0621+1002.}
\label{fig:J0621+1002secspecs}
\end{figure*}

\begin{figure*}
\centering
\includegraphics[width=1.0\textwidth]{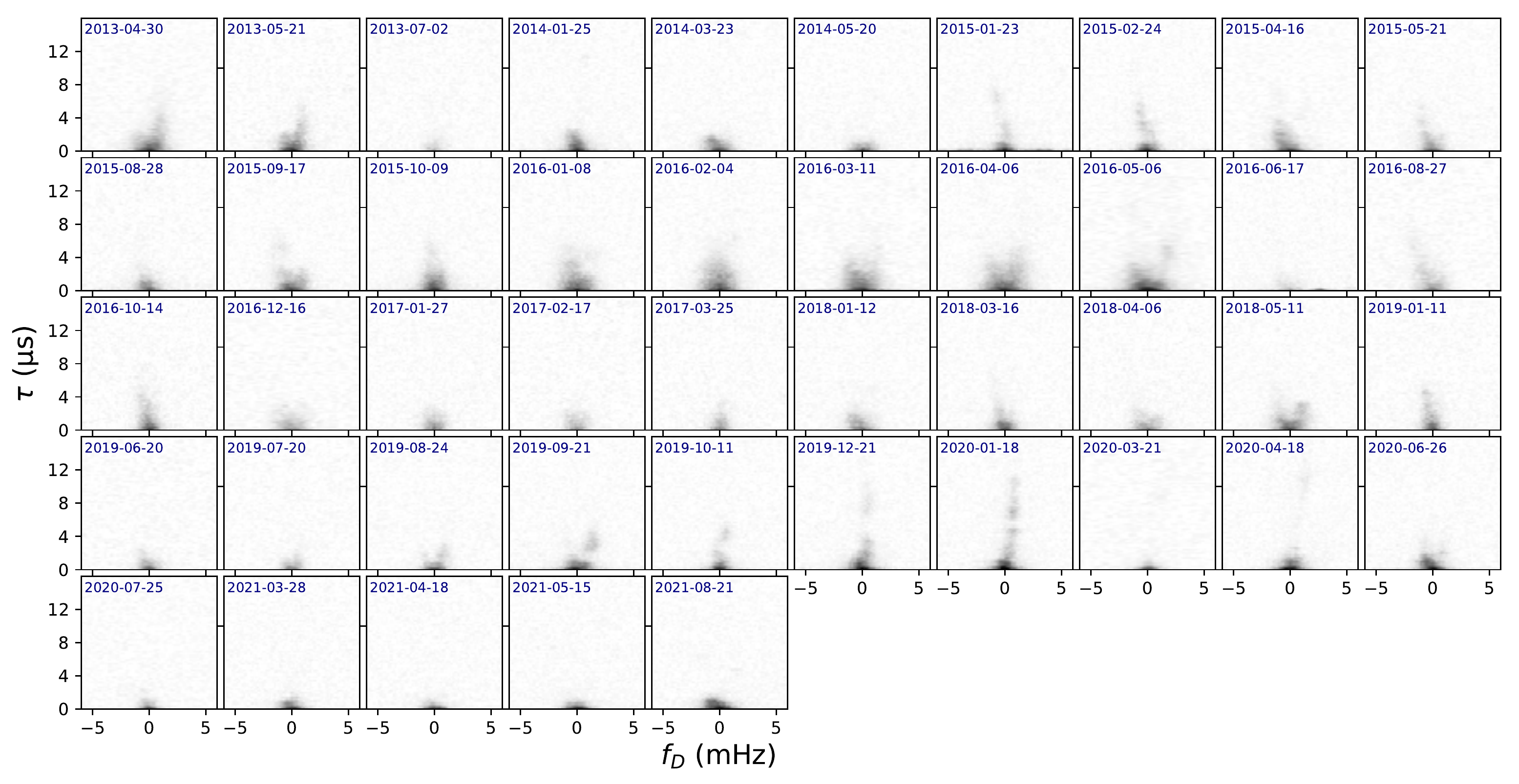}
\vspace{-0.5cm}
\caption{Secondary spectra of all observations of PSR J1600$-$3053.}
\label{fig:J1600-3053secspecs}
\end{figure*}

\begin{figure*}
\centering
\includegraphics[width=1.0\textwidth]{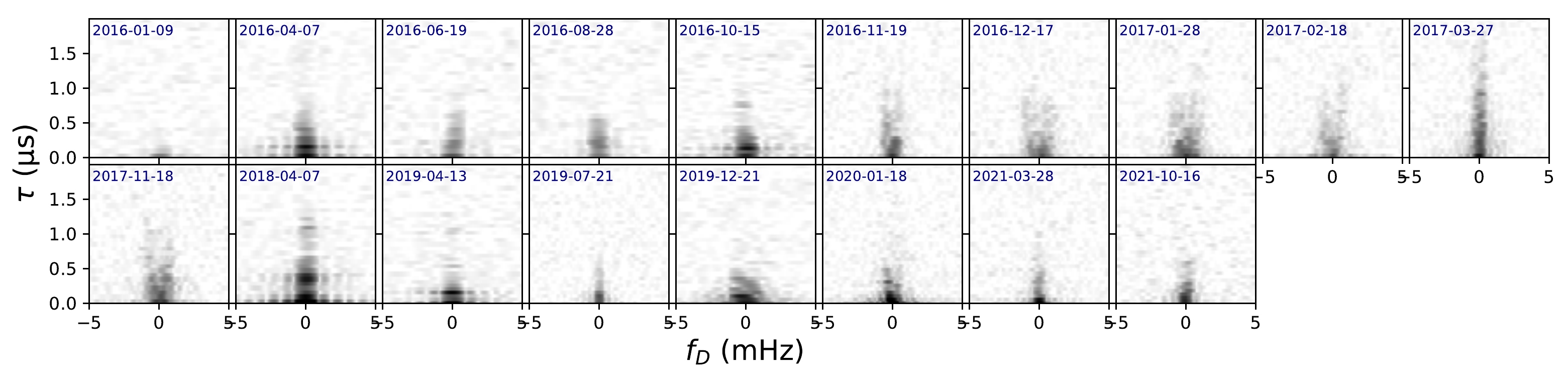}
\vspace{-0.5cm}
\caption{Secondary spectra of all observations of PSR J1918$-$0642.}
\label{fig:J1918-0642secspecs}
\end{figure*}


\begin{figure*}
\centering
\includegraphics[page=1,width=1.0\textwidth]{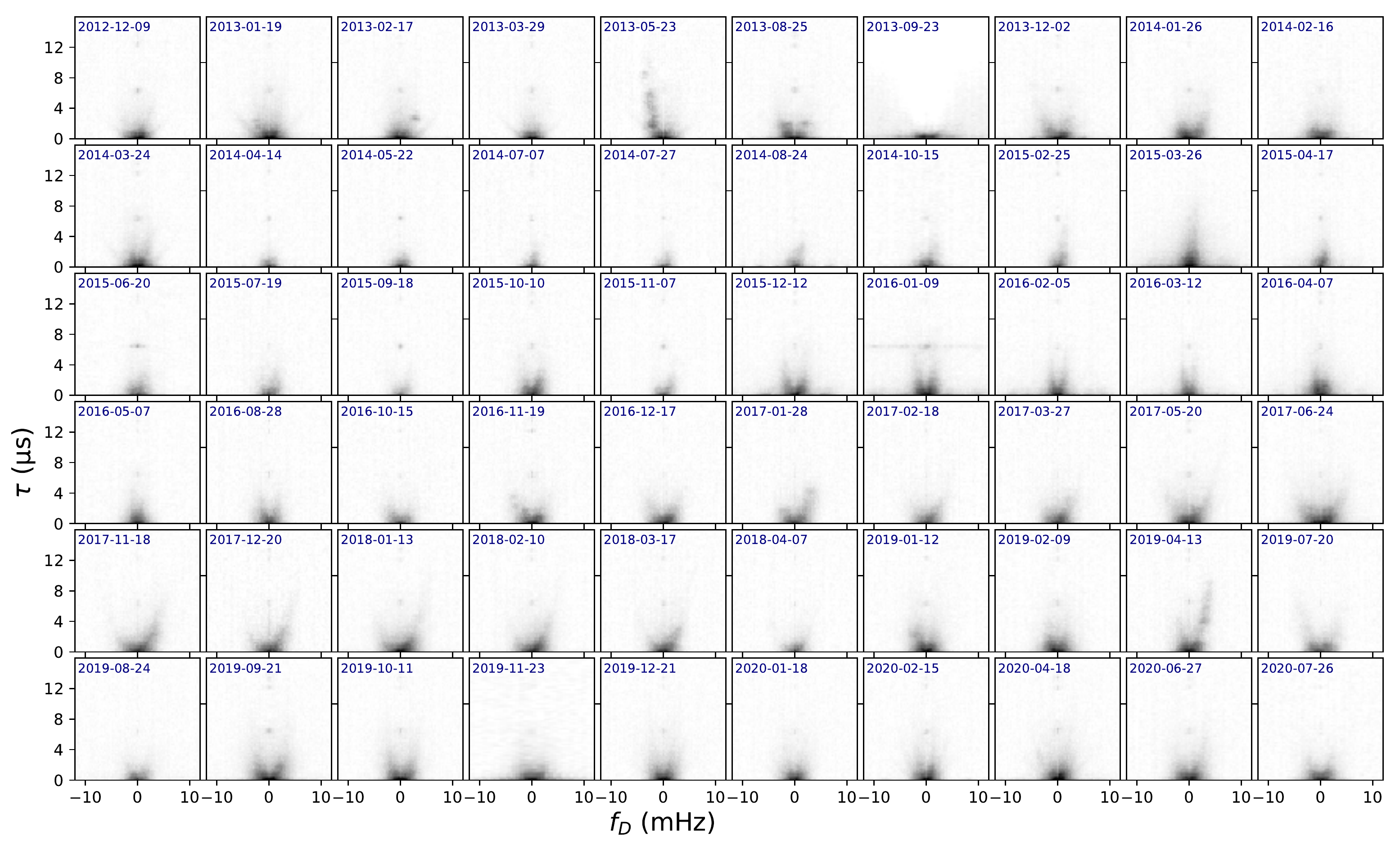}
\vspace{-0.5cm}
\caption{Secondary spectra of all observations of PSR B1937+21.}
\label{fig:B1937+21secspecs}
\end{figure*}

\bsp	
\label{lastpage}
\end{document}